\documentclass[conference]{IEEEtran}
\IEEEoverridecommandlockouts
\usepackage{cite}
\usepackage{amsmath,amssymb,amsfonts}
\usepackage{algorithmic}
\usepackage{graphicx}
\usepackage{textcomp}
\usepackage{xcolor}

\definecolor{mydarkred}{RGB}{180,0,0}

\usepackage{multirow}
\usepackage{bm}
\usepackage{stfloats}
\usepackage{makecell}
\usepackage{soul}
\usepackage{float}
\usepackage{comment}
\usepackage{enumitem}
\usepackage{url}

\def\BibTeX{{\rm B\kern-.05em{\sc i\kern-.025em b}\kern-.08em
    T\kern-.1667em\lower.7ex\hbox{E}\kern-.125emX}}
\begin{document}

\title{Towards Zero Rotation and Beyond: Architecting Neural Networks for Fast Secure Inference with Homomorphic Encryption}


\author{\IEEEauthorblockN{Yifei Cai}
\IEEEauthorblockA{
\textit{Iowa State University}\\
Ames, IA \\
yifeic@iastate.edu}
\and
\IEEEauthorblockN{Yizhou Feng}
\IEEEauthorblockA{
\textit{Old Dominion University}\\
Norfolk, VA \\
yfeng002@odu.edu}
\and
\IEEEauthorblockN{Qiao Zhang}
\IEEEauthorblockA{
\textit{Shandong University}\\
Qingdao, Shandong \\
qiao.zhang@sdu.edu.cn}
\and
\IEEEauthorblockN{Chunsheng Xin}
\IEEEauthorblockA{
\textit{Iowa State University}\\
Ames, IA \\
cxin@iastate.edu}
\and
\IEEEauthorblockN{Hongyi Wu}
\IEEEauthorblockA{
\textit{University of Arizona}\\
Tucson, AZ \\
mhwu@arizona.edu}
\thanks{*This work
has been accepted for publication at the IEEE Conference on Secure and Trustworthy Machine Learning (SaTML). The final version will be available on IEEE Xplore.}
}

\maketitle

\begin{abstract}
Privacy-preserving deep learning addresses privacy concerns in Machine Learning as a Service (MLaaS) using Homomorphic Encryption (HE) for linear computations. Nevertheless, the high computational cost remains a challenge. 
While prior work has attempted to improve the efficiency, most are built upon models originally designed for plaintext inference. These models are inherently limited by architectural inefficiencies when adapted to HE settings.
We argue that substantial efficiency improvements can be achieved by designing networks specifically tailored to the unique computational characteristics of HE, rather than retrofitting existing plaintext models. 
Our design comprises two main components: the building block and the overall architecture. 
The first, StriaBlock, targets the most expensive HE operation—Rotation. It integrates ExRot-Free Convolution and a novel Cross Kernel, completely eliminating the need for external Rotation and requiring only 19\% of the internal Rotation operations compared to plaintext models.
The second component, the architectural principle, includes the Focused Constraint Principle, which limits cost-sensitive factors while preserving flexibility in others, and the Channel Packing-Aware Scaling Principle, which dynamically adapts bottleneck ratios based on ciphertext channel capacity that varies with network depth. These strategies efficiently control the local and overall HE cost, enabling a balanced architecture for HE settings.
The resulting network, StriaNet, is comprehensively evaluated. While prior works primarily focus on small-scale datasets such as CIFAR-10, we conduct an extensive evaluation of StriaNet across datasets of varying scales, including large-scale (ImageNet), medium-scale (Tiny ImageNet), and small-scale (CIFAR-10) benchmarks. At comparable accuracy levels, StriaNet achieves speedups of 9.78$\times$, 6.01$\times$, and 9.24$\times$ on ImageNet, Tiny ImageNet, and CIFAR-10, respectively.
\end{abstract}


\section{Introduction}
\label{introduction}

Deep learning (DL) has become a ubiquitous tool, significantly enhancing productivity and solving complex problems with profound societal implications in the present era. However, the construction of a DL model necessitates access to vast datasets, substantial computational capabilities, and specialized expertise. That is a formidable obstacle often beyond the reach of many users and organizations. In response to this challenge, the concept of Machine Learning as a Service (MLaaS) has emerged as a pragmatic solution~\cite{10.14778/3282495.3282499}.
MLaaS empowers service providers equipped with ample resources to develop well-trained, advanced DL models, offering them as a service accessible to a wide range of users.

However, privacy emerges as a paramount concern in MLaaS, where clients are reluctant to share their valuable private data with the server, and the server guards its model parameters as valuable intellectual property. To mitigate this privacy dilemma, privacy-preserving MLaaS (PP MLaaS) integrates cryptographic primitives into the computation process of DL models, as demonstrated in several recent frameworks, such as Cryptonets~\cite{gilad2016cryptonets}, SecureML~\cite{mohassel2017secureml}, MiniONN~\cite{liu2017oblivious}, GAZELLE~\cite{juvekar2018gazelle}, DELPHI~\cite{244032}, CrypTFlow2~\cite{rathee2020cryptflow2}, Cheetah~\cite{279898}, Iron~\cite{hao2022iron}, BOLT~\cite{pang2023bolt} and BumbleBee~\cite{lu2023bumblebee}.
Commonly adopted cryptographic primitives in these frameworks include Homomorphic Encryption (HE)~\cite{fan2012somewhat}, Garbled Circuits (GC)~\cite{bellare2012foundations}, Oblivious Transfer (OT)~\cite{brassard1986all}, and Secret Sharing (SS)~\cite{shamir1979share}.

Fully Homomorphic Encryption (FHE)~\cite{gentry2009fully,brakerski2014efficient} enables arbitrary computations on encrypted data without decryption or private key access, while Partially Homomorphic Encryption (PHE) supports a limited set of functions, such as additions or depth-bounded operations. PHE is simpler and faster than FHE but less versatile. FHE is often used in scenarios with minimal server-client interaction, requiring nonlinear functions to be approximated as polynomials and relying on computationally expensive bootstrapping to manage noise~\cite{ao2024autofhe}. SS schemes, which involve multiple parties, are commonly applied in settings such as federated learning and distributed computation. OT is typically used to securely exchange messages between parties and facilitate the secure evaluation of nonlinear operations~\cite{brassard1986all}.

\vspace*{0.05in}
\noindent\textbf{Efficiency Challenges for PP MLaaS: }
Given that DL models consist of both linear and nonlinear functions, some frameworks adopt hybrid cryptographic primitives that utilize HE for linear computations and Multi-Party Computation (MPC) protocols—such as GC and OT—for nonlinear operations. State-of-the-art examples include HE-GC-based frameworks like GAZELLE and DELPHI, as well as HE-OT-based frameworks such as CrypTFlow2 and Cheetah. Alternatively, some frameworks~\cite{kim2023hyphen,ao2024autofhe} approximate nonlinear functions, such as ReLU, using polynomials, thereby enabling their evaluation within the HE domain as part of linear computation.
Overall, HE plays a pivotal role in PP MLaaS frameworks, which motivates this work to focus specifically on HE. Many real-time applications demand rapid response times~\cite{AmazonAlexa}; however, the use of cryptographic primitives such as HE introduces significant computational overhead, limiting the practicality of current solutions. For example, performing inference on a single ImageNet image using ResNet-50~\cite{he2016deep} within state-of-the-art privacy-preserving frameworks~\cite{juvekar2018gazelle,244032,rathee2020cryptflow2,zhang2021gala,279898} typically takes approximately 100 to 400 seconds, rendering such latency infeasible for real-world deployment.

\noindent\textbf{Existing Efficiency Enhancement Approaches and Limitations:}
To enhance the efficiency of PP MLaaS, several studies have focused on optimizing inference under HE. Some works, such as SpENCNN~\cite{ran2023spencnn} and HyPHEN~\cite{kim2023hyphen}, explore HE-aware data packing schemes to better exploit SIMD parallelism, achieving approximately 1.85$\times$ speedup. Cheon et al.\cite{cheon2024batch} introduce a channel-by-channel convolution approach whose efficiency scales with batch size, reaching up to 5.2$\times$ speedup with a large batch size of 512. Similarly, HeLayers\cite{aharoni2020helayers} adopts HE-friendly data packing combined with tailored convolution operations, yielding a 1.2$\times$ speedup and further improvements as batch size increases.
Other approaches, such as AutoFHE~\cite{ao2024autofhe} and REDsec~\cite{folkerts2021redsec}, apply function-level optimizations, including polynomial approximations and discretized network representations, to accelerate inference under HE. However, these methods may result in trade-offs in model accuracy. 
Another representative work, Cheetah~\cite{279898}, adopts specialized encoding schemes and polynomial multiplication to avoid the high cost from HE Rotation.
In addition, techniques such as MOSAIC~\cite{cai2024mosaic} and the work by Legiest et al.~\cite{legiest2023neural} emphasize model compression through pruning and quantization to reduce the complexity of the original model  while preserving inference accuracy.

\begin{figure}[t]
\centering
\includegraphics[trim={0cm 0cm 0cm 0cm}, clip, scale=0.252]{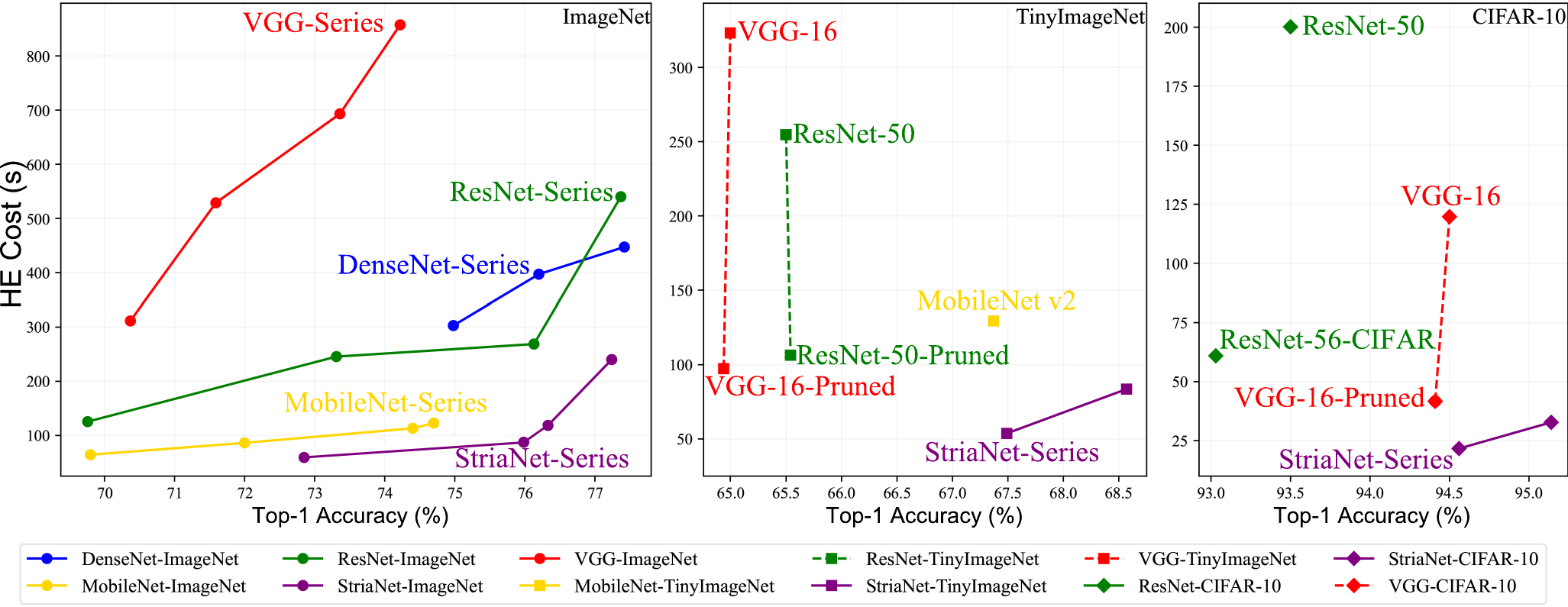}
\vspace*{-0.25in}
\caption{Performance of StriaNet across ImageNet, Tiny ImageNe and CIFAR-10 datasets.}
\vspace*{-0.2in}
\label{curve}
\end{figure}

Despite these advancements, existing methods represent optimizations built upon original models and are inherently constrained by architectures designed for plaintext inference. Models such as VGG and ResNet were originally developed to minimize computational costs in plaintext settings, which can be inefficient—or even detrimental—when applied to HE scenarios. For instance, the pruning-based method MOSAIC~\cite{cai2024mosaic} reduces the inference time of VGG-16 on CIFAR-10 from 119.72s to 41.69s, achieving a 2.87$\times$ speedup (see Figure~\ref{curve}). While this performance gain is notable, substantial room for improvement remains. The core issue is that the baseline model—VGG—is excessively costly in ciphertext settings, and pruning faces inherent limitations, as it must preserve essential structural components to maintain accuracy. \textit{\textbf{But what if the core structures of plaintext models are themselves inherently inefficient for HE?}} In such cases, the effectiveness of any optimization is ultimately bounded by the plaintext-oriented design of the original model architecture.

\vspace*{0.05in}
\noindent\textbf{Reevaluating Plaintext Model Efficiency Under HE:}
For plaintext DL inference, several classic backbone neural networks are widely used, including VGG~\cite{simonyan2014very}, ResNet~\cite{he2016deep}, DenseNet~\cite{huang2017densely}, and MobileNet~\cite{sandler2018mobilenetv2}. These architectures incorporate various optimizations to improve efficiency in plaintext inference, such as advanced downsampling strategies, dimensionality reduction, channel pruning, and structurally efficient designs. Similar optimization techniques are also evident in many other related works~\cite{ronneberger2015u,chollet2017xception,howard2017mobilenets,iandola2016squeezenet,zhang2018shufflenet}.
As shown in Table~\ref{Divergence_Between_FLOPs_HE_Cost}, we evaluate the inference cost divergence of these models under plaintext and HE settings using two metrics: \textit{FLOPs} and \textit{HE cost}. FLOPs (floating point operations) is a standard metric for quantifying the computational cost of plaintext inference. For ciphertext inference under HE, we adopt an aggregated HE cost metric that unifies the overall cost of different HE operations.
Table~\ref{Divergence_Between_FLOPs_HE_Cost} presents three comparison pairs, illustrating that certain architectures exhibit advantages in the plaintext setting but not under HE inference.
First, consider DenseNet. Owing to its refined dimensional adjustments, DenseNet outperforms ResNet in terms of FLOPs. However, this advantage does not hold in HE settings. Its architecture leads to increased HE cost, reversing the efficiency ranking between ResNet and DenseNet.
Second, we observe an amplification effect under HE. Although ResNet achieves substantially lower FLOPs than VGG—even ResNet-101 is more efficient than VGG-11—this advantage diminishes in the HE context. Some structural elements that are beneficial for plaintext inference introduce disproportionately high costs when executed under HE, amplifying inefficiencies. A similar trend is evident in the comparison between MobileNet and ResNet.
These observations underscore the importance of analyzing the distinct cost distribution characteristics of HE. To develop truly HE-efficient architectures, it is crucial to design networks that explicitly consider and minimize HE cost.

\begin{table}[t]
\caption{Efficiency Reversal: Models Efficient in Plaintext Become Inefficient under HE.}
\vspace*{-0.05in}
\label{Divergence_Between_FLOPs_HE_Cost}
\resizebox{84mm}{!}{
\begin{tabular}{c|rclrclrcl}
\Xhline{1.2pt}
\begin{tabular}[c]{@{}c@{}}Model Pair\end{tabular}     & \multicolumn{3}{c}{\begin{tabular}[c]{@{}c@{}}DenseNet-169\\ vs ResNet-50\end{tabular}} & \multicolumn{3}{c}{\begin{tabular}[c]{@{}c@{}}ResNet-101\\ vs VGG-11\end{tabular}} & \multicolumn{3}{c}{\begin{tabular}[c]{@{}c@{}}MobileNet(1.4)\\ vs ResNet-18\end{tabular}} \\ \hline
Plaintext Settings: FLOPs ($\times10^9$)  & \multicolumn{3}{c}{7.07 \bm{$<$} 7.45}                    & \multicolumn{3}{c}{14.86 \bm{$<$} 19.54}                  & \multicolumn{3}{c}{1.23 \bm{$\ll$} 3.59}\\ \hline
HE Settings: HE Cost (Sec)     & \multicolumn{3}{c}{397.07 \bm{$>$} 268.76}                   & \multicolumn{3}{c}{540.09 \bm{$>$} 311.43}                 & \multicolumn{3}{c}{122.88 \bm{$\approx$} 125.53}                  \\ \hline
\begin{tabular}[c]{@{}c@{}}Relative Cost Ratio\\ (Plaintext$\rightarrow$HE Settings)\end{tabular}     & \multicolumn{3}{c}{\begin{tabular}[c]{@{}c@{}}95\% $\rightarrow$ 148\%\\ Reversed\end{tabular}}                   & \multicolumn{3}{c}{\begin{tabular}[c]{@{}c@{}}76\% $\rightarrow$ 173\%\\ Reversed\end{tabular}}                 & \multicolumn{3}{c}{\begin{tabular}[c]{@{}c@{}}34\% $\rightarrow$ 98\%\\ Advantage Lost
\end{tabular}}                  \\ \Xhline{1.2pt}
\end{tabular}
}
\vspace*{-0.2in}
\end{table}

\vspace*{0.05in}\noindent{\bf Our contributions:} 

\noindent This work addresses a critical gap in current research-the lack of neural network architectures explicitly designed for HE settings. We contend that significant efficiency gains can be achieved by developing models specifically tailored to the computational constraints and unique characteristics of HE, rather than repurposing architectures originally designed for plaintext inference.
The main contributions of this work are summarized as follows:

\vspace*{0.05in}\noindent\textbf{(\uppercase\expandafter{\romannumeral1})}
Firstly, we design the building block.
Through analysis, we find that among the three fundamental HE operations—Rotation, Multiplication, and Addition—Rotation, which includes both in-Rot and ex-Rot operations, is the most computationally expensive due to its reliance on the costly key switching operation~\cite{juvekar2018gazelle, rathee2020cryptflow2,279898}. 
Unlike in plaintext networks, where computational costs are more evenly distributed across layers and closely aligned with model parameter counts, the cost of HE operations—particularly Rotation—is heavily concentrated within specific model structures in ciphertext settings.
To address this challenge, we propose the \textbf{StriaBlock}, a novel building block specifically designed to avoid high-cost HE structures and thereby achieve intrinsic HE efficiency. StriaBlock significantly reduces, or even eliminates, both in-Rot and ex-Rot operations within convolutional layers, thereby minimizing the computational overhead associated with Rotation. It incorporates two key design elements:

\vspace*{0.01in}\noindent\textbf{$\bullet$ ExRot-Free Pattern:} We introduce the exRot-Free pattern, an extremely HE-Efficient convolution layer pattern that eliminates the need for the ex-Rot operation.

\vspace*{0.01in}\noindent\textbf{$\bullet$ Cross Kernel:} We introduce the Cross Kernel, maintaining comparable representational capacity while requiring only 19\% of the in-Rot operations compared to traditional kernel.


\vspace*{0.05in}\noindent\textbf{(\uppercase\expandafter{\romannumeral2})}
Secondly, we establish an architectural mechanism aimed at minimizing rotation operations as well. This mechanism is guided by two key principles. First, to minimize the Rotation cost at a specific model layer, it is essential to identify the dominant contributing factor.
In plaintext settings, the computational cost of a layer is typically influenced by multiple factors—such as the input and output dimensions—relatively equally. However, in ciphertext settings, the dominant HE cost is often governed by a single critical factor. For example, in certain layers, the primary HE cost from in-Rot is predominantly determined by the size of the input channel dimension (see discussion in Sec.~\ref{Architecture}).
Thus, our first principle is:

\vspace*{0.01in}\noindent\textbf{$\bullet$ Focused Constraint Principle:} \textit{Constrain cost-sensitive factor while allowing other parameters to vary freely, improving overall network performance without incurring excessive cost. }
\vspace*{0.01in}

Moreover, during local architectural adjustments, we observe that certain layers are highly sensitive to changes in specific dimensions—small increases can trigger sharp surges in HE cost. In contrast, other layers display high tolerance, permitting significant dimensional expansion with minimal cost impact. We investigate these behavioral patterns and identify HE cost fluctuations as a function of network depth. In HE settings, network depth influences the ciphertext’s channel packing capacity thus affects the complexity of HE operations.
Building on these observations, we extend our local adjustment strategy into a global optimization approach across the entire network and propose the second principle:

\vspace*{0.01in}\noindent\textbf{$\bullet$ Channel Packing-Aware Scaling Principle:} \textit{Dynamically adjust the bottleneck ratio of each building block based on ciphertext channel packing capacity, which varies with network depth. This strategy effectively regulates the overall HE cost and supports the development of a balanced, efficient architecture.}
\vspace*{0.01in}

Following these architectural principles, we construct the network using the proposed building blocks, resulting in \textbf{StriaNet}—a neural network and corresponding design methodology for privacy-preserving deep learning.

\vspace*{0.05in}
\noindent\textbf{(\uppercase\expandafter{\romannumeral3})}
To evaluate the performance of StriaNet, comprehensive experiments are conducted. The existing works~\cite{legiest2023neural,ao2024autofhe,folkerts2021redsec,ran2023spencnn,cheon2024batch,kim2023hyphen,aharoni2020helayers,cai2024mosaic} have only considered small-scale neural networks and datasets, such as CIFAR-10. To compare, we perform an extensive evaluation of StriaNet across datasets of varying scales, including large-scale (\textbf{ImageNet}), medium-scale (\textbf{Tiny ImageNet}), and small-scale (\textbf{CIFAR-10}) benchmarks. As shown in Figure~\ref{curve}, StriaNet achieves excellent performance. At comparable accuracy levels, it delivers speedups of \textbf{9.78}\bm{$\times$}, \textbf{6.01}\bm{$\times$}, and \textbf{9.24}\bm{$\times$} on ImageNet, Tiny ImageNet, and CIFAR-10, respectively. The models and codes are available at: \textit{\url{https://github.com/caiyifei2008/StriaNet}}.

\vspace*{0.05in}\noindent\textbf{(\uppercase\expandafter{\romannumeral4})}
Our HE-Efficient architecture offers a potential solution to the communication challenges identified in Cheetah~\cite{279898}. Specifically, our exRot-Free pattern architecture minimizes HE computational cost without introducing additional communication overhead. Compared to Cheetah, StriaNet maintains the same computational complexity while significantly optimizing the communication of the target convolutional layer—achieving a \textbf{31.4}\bm{$\times$} reduction. This leads to a \textbf{13.1}\bm{$\times$} speedup for the target layer and a \textbf{1.73}\bm{$\times$} overall speedup in end-to-end inference.

\vspace*{0.05in}
The remainder of this paper is organized as follows: Sec.\ref{preliminary} introduces the system model, threat model, security analysis, and packed HE. Sec.\ref{HE-Efficient-Network} proposes StriaNet, a highly HE-Efficient DL Network and architecting methodology, including a novel building block-StriaBlock and the HE-Efficient architecture Principles. In Sec.\ref{Evaluation}, we discuss the experimental results. Sec.\ref{conclusion} concludes the paper.

\section{Preliminaries}
\label{preliminary}
\vspace*{-0.05in}

\subsection{System Model}
\label{System_Model}

\begin{figure}[t]
\centering
\includegraphics[trim={0cm 0cm 0cm 0cm}, clip, scale=0.5]{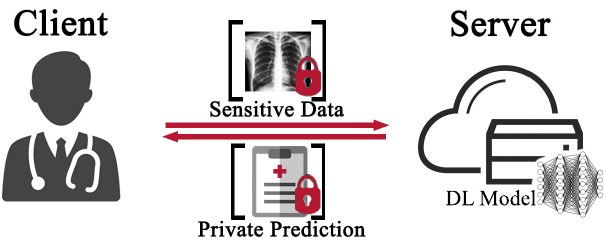}
\vspace*{-0.1in}
\caption{Privacy-preserving MLaaS.}
\vspace*{-0.2in}
\label{MLaaS}
\end{figure}

In this study, we focus on PP MLaaS, depicted in Figure~\ref{MLaaS}. It involves an interaction between two entities: the client ($\mathcal{C}$) with sensitive data, exemplified by medical records, and the server ($\mathcal{S}$) with a high-performance DL model capable of generating prediction. 
Privacy concerns arise during their collaborative exchange, with the client adamant about preventing any disclosure of its private data, including to the server. Simultaneously, the server is reluctant to disclose model parameters due to substantial training resources. 
Consequently, the objective of PP MLaaS is to protect the client's input from the server while maintaining the opacity of the server's model parameters.

Aligned with prior contributions in the domain of PP MLaaS~\cite{gilad2016cryptonets,mohassel2017secureml,liu2017oblivious,juvekar2018gazelle,demmler2015aby,riazi2018chameleon,rouhani2018deepsecure,mohassel2018aby3,wagh2020falcon,patra2020blaze,jiang2018secure,riazi2019xonn,244032,rathee2020cryptflow2,tan2021cryptgpu,wagh2019securenn,zhang2021gala,kumar2020cryptflow,zhang2018gelu,ng2021gforce,boemer2020mp2ml,sureshaby2,279898,dong2022fusion}, our focus centers on deep CNNs, a pivotal and highly successful class of DL models extensively employed across diverse applications~\cite{schroff2015facenet,krizhevsky2012imagenet,simonyan2014very}.
This study aims to improve the computational efficiency of linear operations, which comprise over 90\% of the total inference cost in DL models~\cite{juvekar2018gazelle,rathee2020cryptflow2}. Within CNNs, two core linear operations—convolution and dot product—are critical, with convolution being particularly dominant in terms of computational cost during inference. Experimental evidence demonstrates the significant cost contribution of convolution in PP MLaaS; for instance, it accounts for approximately 99.05\% of the linear computation in ResNet-50~\cite{rathee2020cryptflow2,279898,aharoni2020helayers}. Convolution involves convolving the input feature map with the kernels of the convolutional layer. In the context of MLaaS, the kernels is situated on the server, whereas the input feature is acquired from the client.
Meanwhile, we follow established strategies for privacy-preserving nonlinear operations, such as max-pooling and ReLU activation, in accordance with state-of-the-art frameworks~\cite{juvekar2018gazelle,rathee2020cryptflow2,279898,kim2023hyphen,ao2024autofhe}.

\vspace*{-0.1in}
\subsection{Threat Model and Security Analysis}
\label{Thread_Model}

We adhere to the semi-honest adversary model, a model widely employed in state-of-the-art PP MLaaS frameworks like GAZELLE~\cite{juvekar2018gazelle}, DELPHI~\cite{244032}, CrypTFlow2~\cite{rathee2020cryptflow2}, Cheetah~\cite{279898}, and MiniONN~\cite{liu2017oblivious}. Within this model, both the client and the server conscientiously follow the protocol while attempting to infer additional information from the exchanged messages.
Following the same definition of the cryptographic inference protocol in DELPHI~\cite{244032}, the server holds a model $\mathcal{W}$ consisting of $N$ layers $W_1,W_2,\cdots,W_N$, and the client holds the input vector $x$. During the PP MLaaS interactions, before sending it to the server, the input vector $x$ is encrypted into ciphertext $\mathbf{x}$. On the server side, linear computations, such as convolutions, can be represented as $\mathbf{y}=F(W_i,\mathbf{x})$ where $i \in \{1,2,\cdots,N\}$. The security of this process relies on the semantic security of homomorphic encryption algorithms, such as BFV~\cite{fan2012somewhat} and CKKS~\cite{cheon2017homomorphic}.
StriaNet, in particular, is a highly HE-Efficient DL network architecture that requires fewer HE operations for the same model capacity. From a secure inference standpoint, StriaNet maintains the same level of security, as it introduces no additional computational modules and does not rely on new algorithms. StriaNet is compatible with existing PP MLaaS frameworks and optimizations, making it a strong baseline backbone model for HE-based scenarios. It maximizes the efficiency of these optimizations by offering a better baseline.
Overall, under the semi-honest threat model, StriaNet is considered secure.

\subsection{Packed HE}
\label{Packed_HE}

Homomorphic Encryption (HE) is recognized as a significant breakthrough in the cryptographic domain due to its ability to facilitate linear computations between ciphertext without the need for decryption~\cite{paillier1999public}. This capability enables computational results to be directly calculated in encrypted form, preserving confidentiality.
HE plays a pivotal role in the domain of PP MLaaS, where the client can encrypt sensitive data and send ciphertext to the server. Subsequently, the server can perform computations directly over the ciphertext using its DL model, obtaining prediction results without compromising access to the client's confidential information~\cite{brakerski2013packed,10.14778/3282495.3282499}.


In plaintext deep learning, linear functions are accomplished through common multiply-add operations. However, in the HE-based scenario~\cite{juvekar2018gazelle,rathee2020cryptflow2}, all linear operations, including convolution in convolutional layers and dot product operations in FC layers, are realized through the integration of three fundamental HE operations: 

\begin{itemize}[left=0pt]
\item \textbf{Homomorphic Addition (Add, ``$\oplus$"):} Consider two plaintext vectors, denoted as $\bm{v}_1$ and $\bm{v}_2$, which are respectively packed and encrypted into $[\bm{v}_1]_{\mathcal{C}}$ and $[\bm{v}_2]_{\mathcal{C}}$. Here, $[\cdot]_{\mathcal{C}}$ signifies ciphertext encrypted by the client. The Add operation between ciphertext $[\bm{v}_1]_{\mathcal{C}}$ and $[\bm{v}_2]_{\mathcal{C}}$ mirrors common plaintext addition but operates element-wise, yielding the sum as ciphertext $[\bm{v}_1 \oplus \bm{v}_2]_{\mathcal{C}}$.
\item \textbf{Homomorphic Multiplication (Mult, ``$\odot$"):} The Mult operation between ciphertext $[\bm{v}_1]_{\mathcal{C}}$ and plaintext vector $\bm{v}_2$ is executed through element-wise multiplication, resulting in the product ciphertext $[\bm{v}_1 \odot \bm{v}_2]_{\mathcal{C}}$. It should be noted that, in typical CNN inference, there is no scenario where a Mult operation occurs between two ciphertexts.
\item \textbf{Homomorphic Rotation (Rot):} The rotation operation performs a cyclic shift of the elements within a ciphertext. Given a ciphertext $[\bm{v}]_{\mathcal{C}}$, rotating it by $i$ positions entails shifting all elements in a cyclic manner by $i$ positions in reverse order. This results in a new ciphertext, denoted as $[\bm{v}{(+i)}]_{\mathcal{C}}$. 
Among the three fundamental HE operations, rotation is the most computationally expensive, as it involves costly key switching procedures~\cite{juvekar2018gazelle,rathee2020cryptflow2,279898}.
\end{itemize}

\noindent It's essential to note that both the Add and Mult operations function in an \textbf{element-wise manner}. Thus, directly summing up the values of a vector is not feasible, as these elements reside in different slots. The \textbf{Rotation} operation addresses this challenge by rotating the ciphertext, aligning internal elements for subsequent operations.

\vspace*{0.05in}
Within PP MLaaS frameworks, it is common practice to encrypt and aggregate multiple input channels into a single ciphertext to enhance efficiency~\cite{albrecht2021homomorphic,natarajan2021seal,riazi2019xonn}. Packed HE significantly boosts performance by enabling SIMD-style parallelism, where a single ciphertext encapsulates multiple plaintext values. This allows a single operation to process many data points simultaneously, resulting in more than an order-of-magnitude speedup~\cite{juvekar2018gazelle,rathee2020cryptflow2,279898}.
Notably, the size of input features typically diminishes as one progresses from the initial layer to deeper layers within a DL model due to the down-sampling. 
For instance, in experiments with the ResNet on the ImageNet dataset, the input size reduces from $224\times224$ at the first layer to $7\times7$ at the deep layer. 
To ensure both safety and efficiency, the length of a single ciphertext is conventionally established at 8192 slots in PP MLaaS frameworks~\cite{albrecht2021homomorphic,natarajan2021seal,riazi2019xonn,juvekar2018gazelle,rathee2020cryptflow2}. As a result, the number of input channels that can be packed within a single ciphertext—referred to as the \textbf{channel packing capacity} and denoted by $\bm{c_n}$—increases significantly, from 2 in shallow layers to 128 in deeper layers. The notation used in this paper is as follows:

\vspace*{-0.05in}
\begin{table}[h]
\centering
\label{symbols}
\resizebox{85mm}{!}{
\begin{tabular}{clcl}
\Xhline{1.2pt}
Notation & Definition & Notation & Definition \\ \hline
$c_i$      & input channels          & $c_o$      & output channels          \\
$k_w$      & width of the kernel          & $k_h$      & height of the kernel          \\
$W$      & width of the input          & $H$      & height of the input          \\
$\mathbb{N}_x$      & the set\{0, 1, \ldots, x-1\}          & $\lceil x \rceil$      & ceiling operation on x          \\
$c_n$      & \multicolumn{3}{l}{channel packing capacity: number of packed channels in a ciphertext}          \\
\Xhline{1.2pt}
\end{tabular}}
\end{table}
\vspace*{-0.15in}

\section{StriaNet: HE-Efficient Deep Learning Network and Architecting Methodology}
\label{HE-Efficient-Network}

To address a critical gap in current research, the lack of neural network architectures explicitly designed for HE settings, we introduce StriaNet, a neural network and accompanying architectural methodology that fully considers the unique computational characteristics of HE-based convolution. StriaNet demonstrates significantly higher HE efficiency compared to existing models originally designed for plaintext inference.

As with all network architecture designs, StriaNet consists of two key components: the \textbf{building block design} and the \textbf{network architectural principle}.
The first component, StriaBlock, incorporates ExRot-Free Convolution and a novel Cross Kernel, eliminating the need for ex-Rot and requiring only 19\% of the in-Rot operations compared to plaintext models.
The second component, the architectural principle, includes Focused Constraint Principle, which limits cost-sensitive factors while preserving flexibility in others, and Channel Packing-Aware Scaling Principle, dynamically adapts bottleneck ratios based on ciphertext channel capacity that varies with network depth. These strategies efficiently control the local and overall HE cost, enabling a balanced architecture for HE settings.

\subsection{Building Block Design}
\label{Building_Block}

\subsubsection{\textbf{exRot-Free Pattern in MIMO}}
\label{Rotation-Free}

\begin{figure}[t]
\centering
\includegraphics[trim={0cm 0cm 0cm 0cm}, clip, scale=0.44]{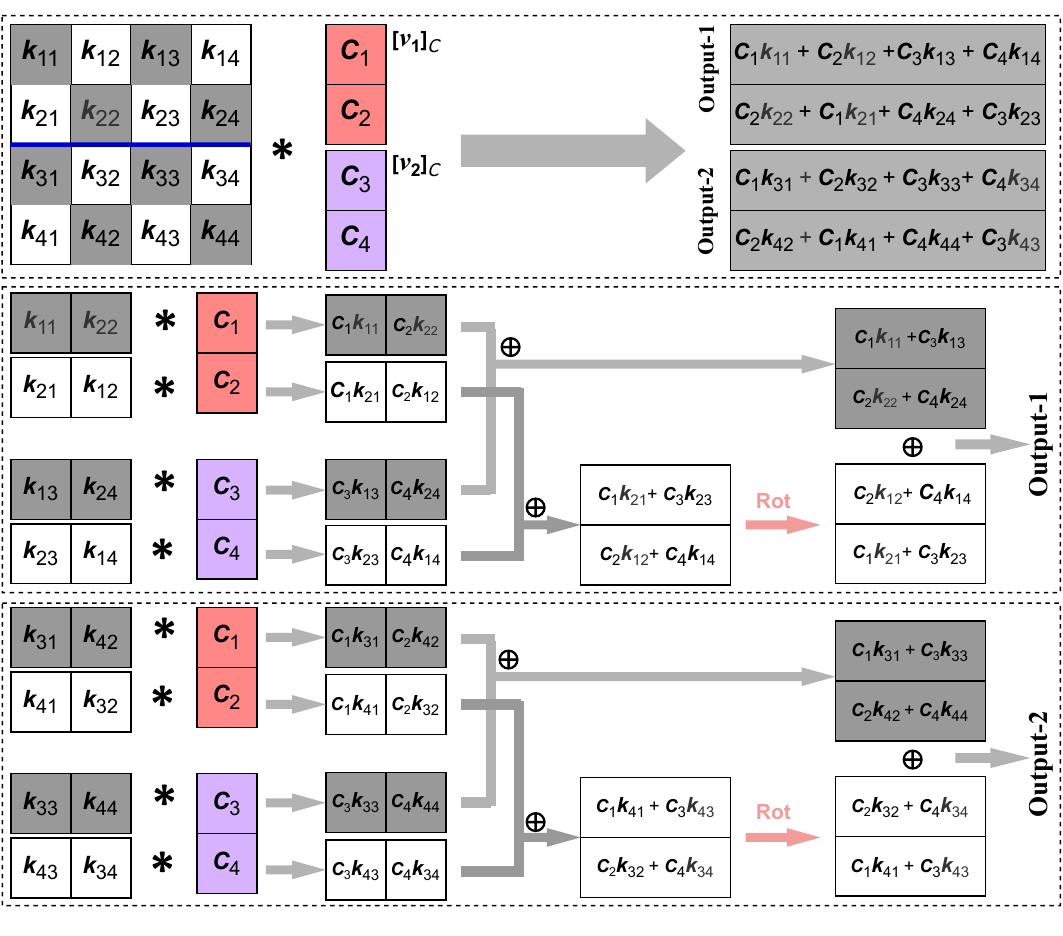}
\vspace*{-0.15in}
\caption{Grouped Out-Rot MIMO ($c_i=c_o=4,c_n =2$).}
\label{CrypTFlow2_Output-Ro}
\vspace*{-0.2in}
\end{figure}

The general convolution computation with Multiple Input Multiple Output (MIMO) is a configuration commonly used in many DL models. 
Figure~\ref{CrypTFlow2_Output-Ro} illustrates the MIMO scheme\footnote{For conciseness, we present the Grouped Output Rotation MIMO scheme, which serves as the default approach in PP MLaaS frameworks~\cite{rathee2020cryptflow2}. For details on an alternative, the Input Rotation MIMO, please refer to Appendix-\ref{moreMIMO}. \textit{Note that the design proposed in this work is applicable to both the Output Rotation and Input Rotation MIMO schemes.}} when $c_n = 2$. In this case, 4 input channels, denoted as $\bm{c}_1, \bm{c}_2, \bm{c}_3, \bm{c}_4$, are packed and encrypted into 2 ciphertext, $[\bm{v_1}]_{\mathcal{C}}$ and $[\bm{v_2}]_{\mathcal{C}}$. These ciphertext are then convolved with 4 filters, four rows of the kernel matrix. Each filter comprises 4 kernels, corresponding to the 4 input channels. Consequently, the output is obtained, consisting of two encrypted ciphertexts, namely Output-1 and Output-2. 
To compute the Output, we first perform a convolution (denoted by the $\ast$ symbol) between the input ciphertext $[\bm{v_1}]_{\mathcal{C}}$ and the kernel group $\{\bm{k}_{11}, \bm{k}_{22}\}$, resulting in the ciphertext $[\bm{c}_1\bm{k}_{11}, \bm{c}_2\bm{k}_{22}]_{\mathcal{C}}$, referred to as the \textit{intermediate ciphertext}. Next, the same input ciphertext $[\bm{v_1}]_{\mathcal{C}}$ is convolved with the kernel group $\{\bm{k}_{21}, \bm{k}_{12}\}$ to produce another intermediate ciphertext $[\bm{c}_1\bm{k}_{21}, \bm{c}_2\bm{k}_{12}]_{\mathcal{C}}$. These intermediate ciphertexts are subsequently combined through a series of additions using the HE Add operation to generate the final Output. To conform with the element-wise computation manner of Packed HE, several HE Rotation operations are applied to realign the intermediate ciphertexts. In the MIMO configuration, this Rotation operation is referred to as external Rotation, abbreviated as \textbf{ex-Rot}.
As discussed in Sec.\ref{Packed_HE}, in HE settings, there are three fundamental HE operations: Rotation, Mult and Add. And among these, Rotation is by far the most computationally expensive operation\cite{juvekar2018gazelle,rathee2020cryptflow2,279898}.
Moreover, based on our analysis, such as ResNet-50 running on the ImageNet dataset, we find that ex-Rot account for a substantial 82.09\% of the total Rotation cost. This observation strongly motivates our design focus: \textit{to prioritize the reduction, and ideally the elimination, of ex-Rot operations.}

\begin{figure*}[t]
\centering
\includegraphics[trim={0cm 0cm 0cm 0cm}, clip, scale=0.27]{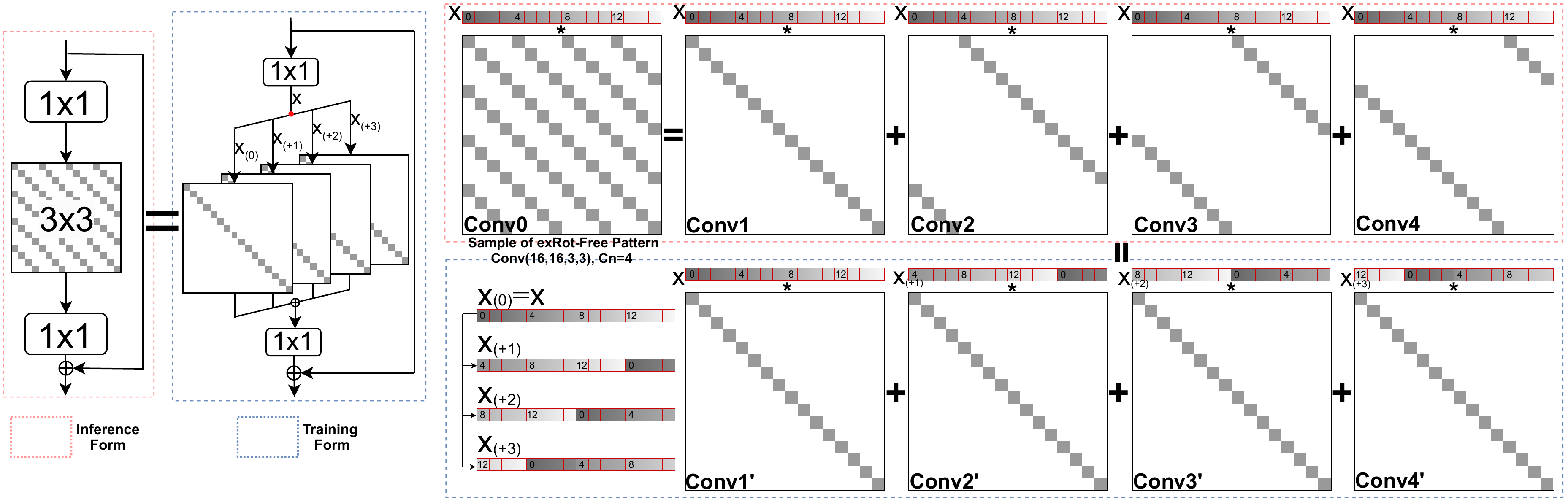}
\vspace*{-0.15in}
\caption{StriaBlock with exRot-Free Pattern ($c_i=c_o=16, c_n =4$).}
\vspace*{-0.2in}
\label{exRot_Free_pattern}
\end{figure*}

\vspace*{-0.1in}
\begin{table}[h]
\renewcommand{\arraystretch}{1.1}
\centering
\resizebox{88mm}{!}{
\begin{tabular}{|l|}
\hline
\multicolumn{0}{|c|}{\textbf{exRot-Free Pattern}} \\
For the kernel with kernel matrix index $[I_{row},I_{col}]$: \\
$I_{row} =  k+c_n\times i$   \quad \text{s.t.} \quad $I_{row}\in \mathbb{N}_{c_o}$,   \\
$I_{col} =  k+c_n\times j$   \quad \text{s.t.} \quad $I_{col}\in \mathbb{N}_{c_i}$,   \\ 
where $k\in \mathbb{N}_{c_n}$; $i \in \mathbb{N}_{m} \leftarrow m=\lceil \frac{c_o}{c_n} \rceil$; $j \in \mathbb{N}_{n} \leftarrow n=\lceil \frac{c_i}{c_n} \rceil.$                                \\ \hline
\end{tabular}}
\vspace*{-0.1in}
\end{table}

Upon deep analysis of the MIMO schemes, we observe that not all convolutional layer structures necessitate ex-Rot. Specifically, only those structures involving intermediate ciphertexts whose channel order differs from that of the original input require ex-Rot for reordering. Therefore, the need for ex-Rot can be eliminated if the convolutional layer is constructed exclusively using structures (i.e., convolution kernels) associated with intermediate ciphertexts whose channel order aligns with the original input.
Building on this insight, we design a universally \textbf{exRot-Free pattern}. This pattern is characterized by the main diagonal of each $c_n \times c_n$ kernel matrix, which is visually highlighted in grey in Figure~\ref{CrypTFlow2_Output-Ro}. Kernels adhering to this exRot-Free pattern do not require any ex-Rot operations. Moreover, layers composed entirely of this structure also eliminate the need for ex-Rot during the MIMO process. (Note that $c_n$, the channel packing capacity, is one central element that permeates our design and will be further discussed in subsequent sections.)

Figure~\ref{exRot_Free_pattern} illustrates the building block that adopts the exRot-free pattern, exhibiting a distinctive striated texture. We name this block \textbf{StriaBlock}. StriaBlock is designed as an inverted residual block composed of a sequence of three convolutional layers with kernel sizes of $1\times1$, $3\times3$, and $1\times1$, respectively. The $3\times3$ layer serves as the primary feature extractor, while the two $1\times1$ layers are responsible for channel scaling and mixing. As shown in Figure~\ref{exRot_Free_pattern}, the $3\times3$ layer adopts the exRot-Free pattern, which brings two major benefits.

First, from the perspective of \textbf{computational efficiency}, the $3\times3$ layer is typically the most computation-intensive part of the block, particularly due to its heavy reliance on ex-Rot operations. Applying the exRot-Free pattern to this layer maximizes the efficiency gain, reducing the overall HE cost by up to 60\%.

Second, in terms of \textbf{network capacity}, a layer following the exRot-Free pattern can, in principle, be decomposed into multiple simple depthwise convolutional layers, each operating on input data with a different channel order. As illustrated on the right side of Figure~\ref{exRot_Free_pattern}, the top row shows a convolutional layer (Conv0) following the exRot-Free pattern convolving with the input $x_{(0)}$. This corresponds to the inference form. Conv0 can be deconstructed into four sub-layers (Conv1 to Conv4), each also convolving with the original input $x_{(0)}$.
These sub-layers can then be transformed by reordering their input channels (i.e., columns of the kernel matrix) by 4, 8, and 12 channels, resulting in Conv1' to Conv4'. Simultaneously, the input $x_{(0)}$ is reordered accordingly into $x_{(+1)}$, $x_{(+2)}$, and $x_{(+3)}$, forming the training form.
Importantly, the training and inference forms are functionally equivalent. During training on the server side, all operations are conducted in plaintext, making channel reordering straightforward. After training, the model is reassembled into Conv0 with the original input $x_{(0)}$, requiring no HE Rotation operation during encrypted inference.

Functionally, the exRot-Free pattern behaves like multiple parallel depthwise convolutions on inputs with varied channel orders. This structure enables effective extraction of diverse features from the input, which are then integrated and preserved by the subsequent $1\times1$ layer. As a result, StriaBlock offers both high HE efficiency and strong representational capacity, making it a powerful and practical building block for secure inference with HE.

\subsubsection{\textbf{Cross Kernel in SISO}}
\label{Cross_Kernel}
The proposed exRot-Free pattern eliminates the need for ex-Rot operations. Furthermore, we continue to address the other HE Rotation operation, in-Rot, which is required during the inner kernel computation, the convolution within Single Input and Single Output (SISO). 

\begin{figure}[t]
\centering
\includegraphics[trim={0cm 0cm 0cm 0cm}, clip, scale=0.5]{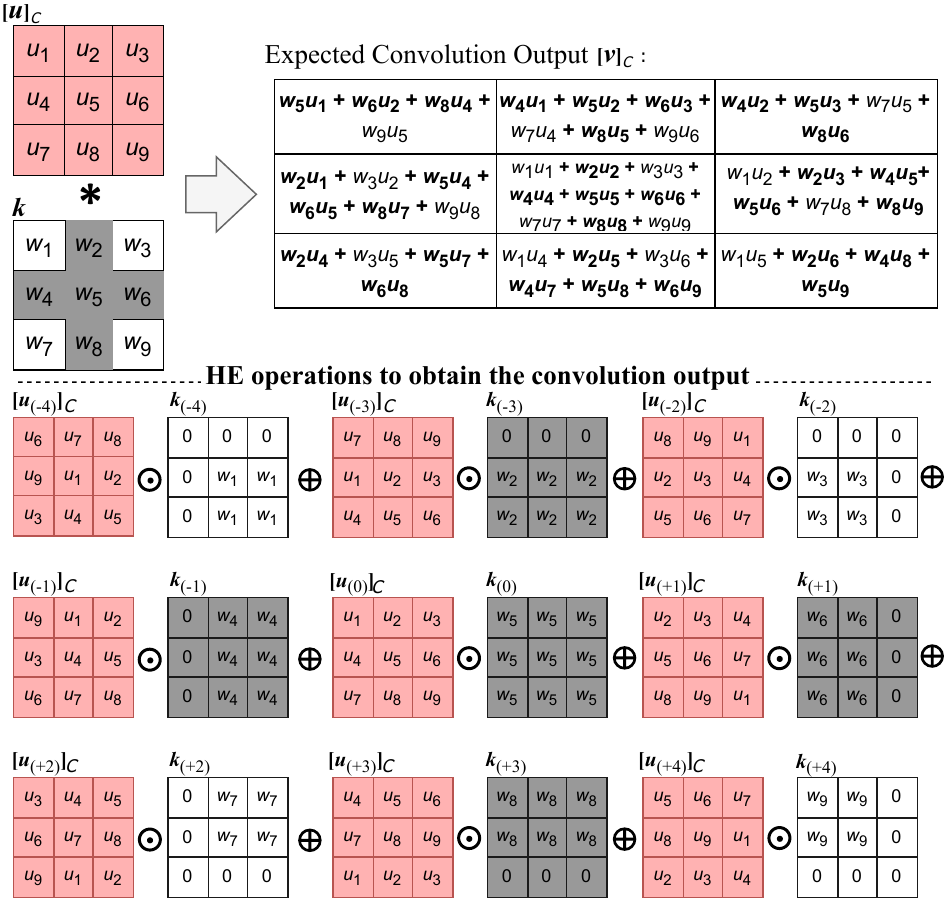}
\vspace*{-0.15in}
\caption{SISO ($W=H=3,k_w=k_h=3$).}
\label{SISO_Cross}
\vspace*{-0.23in}
\end{figure}

In Figure~\ref{SISO_Cross}, the client encrypts its input $\bm{u}$ (with the size of $W\times H$) as $[\bm{u}]_{\mathcal{C}}$. This ciphertext is then sent to the server to convolve with the corresponding plaintext kernels $\bm{k}$ (with the size $k_w\times{k_h}$, and the $*$ in the figure denotes the convolution operator). The encrypted convolution output $[\bm{v}]_{\mathcal{C}}$ is obtained.
Specifically, the convolution operation is achieved by first placing the kernel at each location of the input feature and then summing up all the element-wise products between the input values and the corresponding kernel values covering the respective positions. 
However, as previously discussed, the server cannot directly perform the summation because, in the HE-based setting, the Add operation functions in an element-wise manner. Specifically, the Add operation can only sum values located at the same slot across multiple ciphertexts; it cannot aggregate values from different slots within a single ciphertext. To address this limitation, the Rotation operation is employed to shift the ciphertext, aligning the desired slots for correct summation.

As illustrated in Figure~\ref{SISO_Cross}, to conform with the element-wise computation manner, the original inputs $[\bm{u}_{(0)}]_{\mathcal{C}}$ must be rotated to $[\bm{u}_{(-4)}]_{\mathcal{C}}\sim[\bm{u}_{(+4)}]_{\mathcal{C}}$ using eight in-Rot operations. This introduces a significant computational cost, especially considering the large-scale input channels in the overall network. Even with the use of hoisting techniques~\cite{juvekar2018gazelle}, which can improve the batch efficiency of such operations, the cost remains substantial.
According to our analysis, not all parameters within a kernel incur the same cost. For instance, in a $3\times3$ kernel, every parameter except the central one, $k_5$, requires an in-Rot operation. However, the cost associated with these operations can vary significantly. In typical settings, for efficiency and security, the length of the HE ciphertext is configured as a power of two (e.g., 8192), and the input dimensions are padded accordingly to a power-of-two size. Let us denote the padded input dimension as $N$. The required in-Rot operations for the kernel weights involve position shifts of $\pm1$, $\pm N$, and $\pm N \pm 1$.
Due to the inherent properties of HE ciphertexts, rotations by positions that are powers of two are generally more efficient~\cite{albrecht2021homomorphic,natarajan2021seal,sealcrypto}. 
Consequently, in-Rot operations with position shifts of $\pm N$ benefit from this efficiency, and those with small shifts such as $\pm 1$ also incur relatively low overhead. In contrast, in-Rot operations by positions $\pm N \pm 1$ are considerably more expensive.


\begin{table}[h]
\vspace*{-0.15in}
\centering
\caption{Kernel in-Rot Cost Analysis.}
\vspace*{-0.05in}
\label{Corner_in-Rot}
\resizebox{80mm}{!}{
\begin{tabular}{c|ccccc}
\Xhline{1.5pt}
Input dimension $N$   &  \multicolumn{1}{c}{4} &  \multicolumn{1}{c}{8}   &  \multicolumn{1}{c}{16}  &  32  &  64\\ \hline
Cross kernel in-Rot (ms) &  23.27 &  22.78  &  \multicolumn{1}{l}{23.8}  &  23.93  &  \multicolumn{1}{c}{23.94}  \\ \hline
Regular kernel in-Rot (ms) &  70.58 &  77.41  &  \multicolumn{1}{l}{91.85}  &  105.00  &  \multicolumn{1}{c}{128.85 }  \\ \hline
Relative cost (\%)  &  \multicolumn{1}{c}{33\%}  &  29\%  &  26\%  &  \multicolumn{1}{c}{23\%}  &  \multicolumn{1}{c}{19\%}  \\ 
\Xhline{1.5pt}
\end{tabular}}
\vspace*{-0.05in}
\end{table}

\begin{figure}[t]
\vspace*{-0.0in}
\centering
\includegraphics[trim={0cm 0cm 0cm 0cm}, clip, scale=0.5]{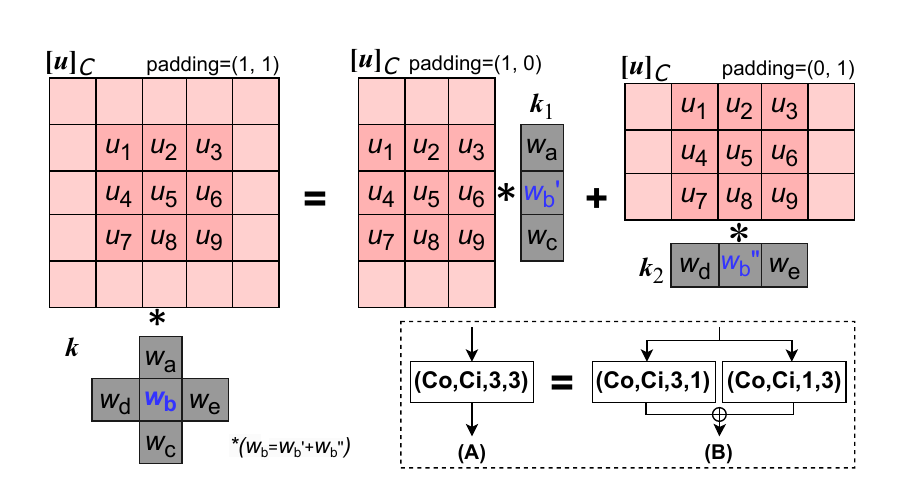}
\vspace*{-0.15in}
\caption{Cross Kernel.}
\vspace*{-0.25in}
\label{cross_kernel}
\end{figure}

We refer to these high-cost model weights as corner weights, as they are typically located at the four corners of a kernel. Then we propose the \textbf{Cross Kernel}, as illustrated in Figure~\ref{cross_kernel}. As the name suggests, the Cross Kernel retains only the five weights along the cross-shaped backbone of the kernel, thereby excluding all expensive corner weights.
As shown in Table~\ref{Corner_in-Rot}, the in-Rot cost of the Cross Kernel is significantly lower than that of the regular kernel, and this advantage becomes increasingly pronounced as $N$ grows. Specifically, when $N = 64$, using the Cross Kernel reduces the in-Rot cost to as low as 19\% of that incurred by a regular kernel.
Importantly, the Cross Kernel fundamentally differs from pruning methods, where model weights are removed post-training. Such pruning typically introduces convergence challenges, including accuracy degradation, and necessitates iterative pruning and fine-tuning, which are both time-consuming and computationally expensive. In contrast, the Cross Kernel is designed without corner weights from the outset, thus avoiding these issues entirely.

This design is achieved through the degenerated Spatially Separable Convolution~\cite{szegedy2016rethinking}. As shown in Figure~\ref{cross_kernel}, the convolution between the input and a Cross Kernel can be equivalent to the sum of two convolutions between the input and two sub-kernels with dimensions $(3,1)$ and $(1,3)$ separately. Thus, one convolutional layer with the Cross Kernel is equivalent to two parallel convolutional layers with $(3,1)$ and $(1,3)$ kernels, respectively.
More generally, for an arbitrary $M \times N$ kernel (where $M$ and $N$ are either odd or adjusted to be odd), the Cross Kernel includes only the weights along the cross backbone—that is, those in the $\frac{M+1}{2}$-th row and $\frac{N+1}{2}$-th column.

A natural question arises: does the Cross Kernel have sufficient representational capacity to be used in practice? First, prior pruning studies~\cite{cai2022hunter,chen2021only} have shown that corner weights are often pruned, suggesting they are not critical to model function. Second, we empirically evaluate the Cross Kernel by replacing all standard kernels with Cross Kernels, and observe at most a 0.32\% minimal impact on accuracy—demonstrating that the Cross Kernel retains sufficient feature extraction capability for practical use.



\subsubsection{\textbf{StriaBlock}}
\label{StriaBlock}
We now consolidate the proposed components. Table~\ref{HEComplexity} compares the HE complexity of a standard convolutional layer with that of StriaNet convolutional layer, which incorporates proposed both the exRot-Free pattern and the Cross Kernel design. The StriaNet convolutional layer offers substantial advantages by completely eliminating the need for ex-Rot operations and significantly reducing the in-Rot operations.
In addition to Rotation, other HE operations include Mult and Add. As shown in the table, the complexity of Mult inherently benefits from optimizations applied to Rotation. Moreover, the Add operation has a minimal impact on the overall cost—contributing as little as 0.5\%~\cite{rathee2020cryptflow2,279898}—and is thus considered negligible in our analysis.

\vspace*{-0.15in}
\begin{table}[h]
\centering
\caption{Complexity Comparison.}
\vspace*{-0.05in}
\label{ComplexityComparison}
\resizebox{85mm}{!}{
\begin{tabular}{c|ccc}
\Xhline{1.5pt}
& \#in-Rot & \#ex-Rot (2 MIMO schemes) & \#Mult \\ \hline
Vanilla Conv  & $\frac{c_{i}(k_{w}k_{h}-1)}{c_{n}}$       & \begin{tabular}[c]{@{}l@{}}$\frac{c_{o}(c_{n}-1)}{c_{n}}$ or $\frac{c_{i}(c_{n}-1)k_{w}k_{h}}{c_{n}}$\end{tabular}       & $\frac{c_{i}c_{o}k_{w}k_{h}}{c_{n}}$    \\ \hline
\textbf{StriaNet Conv} & $\frac{c_{i}(\bm{k_{w}+k_{h}}-1)}{c_{n}}$       & \textbf{0}       & $\frac{c_{i}c_{o}k_{w}k_{h}}{\bm{c_{n}^2}}$    \\ 
\Xhline{1.5pt}
\end{tabular}}
\label{HEComplexity}
\vspace*{-0.1in}
\end{table}

\begin{figure}[t]
\centering
\includegraphics[trim={0cm 0cm 0cm 0cm}, clip, scale=0.272]{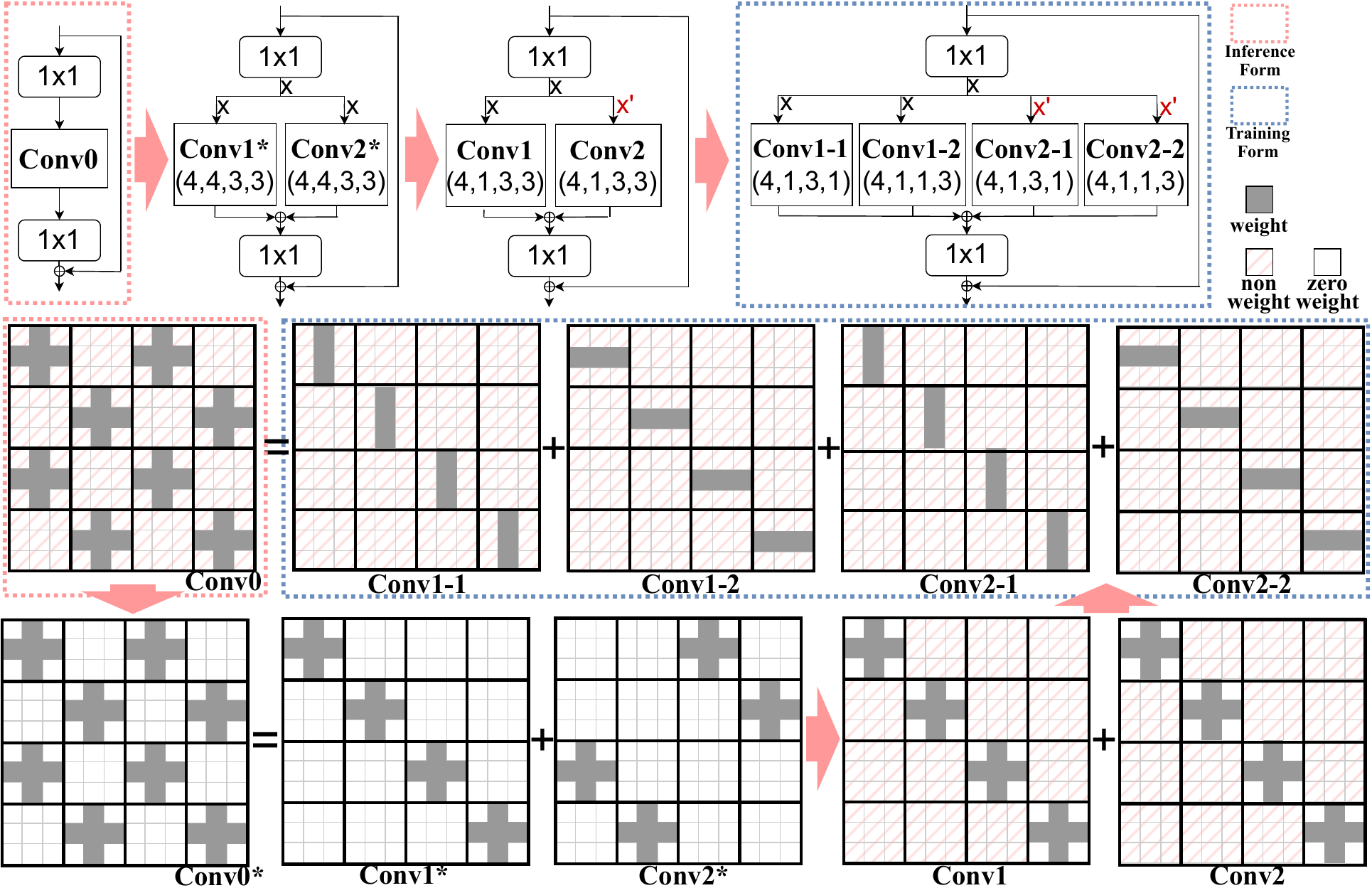}
\vspace*{-0.2in}
\caption{StriaBlock Structural Details. Joint exRot-Free Pattern and Cross Kernel ($c_i=c_o=4,k_w=k_h=3, c_n =2$). ``$=$'' indicates mathematical equivalence and ``$\Rightarrow$'' denotes functional transformation equivalence.}
\vspace*{-0.2in}
\label{StriaBlock_Structure}
\end{figure}

Next, we explore the detailed structure of the StriaBlock using a concise example, as illustrated in Figure~\ref{StriaBlock_Structure}. Conv0 is a convolutional layer comprising four input and four output channels, constructed with $3\times3$ kernels. It incorporates both the exRot-Free pattern and the Cross Kernel design, representing the \textit{Inference Form} that is ultimately deployed on the server. In the figure, grey-highlighted cells denote active weights, while red-hatched cells indicate non-weight positions—these elements are excluded from the model architecture and have no associated parameters.
To deconstruct Conv0 and transition it from the \textit{Inference Form} to a training-compatible form, we conceptually outline a three-step transformation process:

\noindent\textbf{$\bullet$ Conv0$\Rightarrow$ Conv0*= Conv1*+Conv2*:}
In the first step, all non-weight cells in Conv0 are replaced with zero-weight cells, yielding Conv0*. This introduces zero-valued weights in previously unused positions without affecting the computational behavior or functionality of the structure. Conv0* can then be decomposed into two components: Conv1* and Conv2*.

\noindent\textbf{$\bullet$ Conv1*+Conv2*$\Rightarrow$Conv1+Conv2:}
In the second step, the input channel order for Conv2* is rearranged according to the manner as illustrated in Figure~\ref{exRot_Free_pattern}, transforming the input from \textbf{$x$} to \textbf{$x'$}. Additionally, all zero-weight kernels are replaced with non-weight kernels. This results in Conv1 and Conv2, which are two Depthwise Separable convolutional layers.

\noindent\textbf{$\bullet$ Conv1+Conv2$\Rightarrow$Conv1-1+Conv1-2+Conv2-1+Conv2-2:}
In the final step, following the methodology shown in Figure~\ref{cross_kernel}, each Cross Kernel is decomposed into two sub-kernels. This transforms the combined structure (Conv1 + Conv2) into four distinct convolutional layers: Conv1-1, Conv1-2, Conv2-1, and Conv2-2. This configuration represents the \textit{Training Form}, which is used during model training. Importantly, this form is fully equivalent to the original \textit{Inference Form} (Conv0), ensuring a seamless and lossless transition between training and inference.

\vspace*{-0.1in}
\subsection{HE-Efficient Architectural Principle}
\label{Architecture}

In this section, we present the architectural principles of StriaNet. Unlike plaintext model architectures, which are not well-suited for HE, our design is specifically crafted to accommodate the unique computational characteristics of DL models in HE setting.
We begin with a comprehensive analysis of the computation patterns in HE-based DL models to establish the fundamental relationship between HE operation complexity and key architectural factors. Subsequently, we evaluate full DL models within the HE context, with particular attention to a critical variable: the ciphertext’s channel packing capacity, $c_n$, which varies with network depth.

\subsubsection{\textbf{Network Architecture HE Complexity Analysis}}
\label{Dimension}

In plaintext scenarios, convolution is performed using a series of multiply-add operations. The computational complexity in the plaintext domain is quantified in terms of FLOPs, as illustrated in the following equations:
\vspace*{-0.05in}
\begin{equation}
\begin{split}
FLOPs &=  (2c_{i} \times k_{w}k_{h} -1) \times HW \times c_{o},\\
      &\approx  2k_{w}k_{h} \times c_{i}c_{o} \times HW,
\end{split}
\label{FLOPs}
\end{equation}

\vspace*{-0.05in}
\noindent which exhibits one characteristic: \textit{The plaintext cost of a convolutional layer is positively correlated with multiple factors}, including $c_i$ (input channels), $c_o$ (output channels), $k_w$ (kernel width), $k_h$ (kernel height), and $HW$ (feature map size), with each factor contributing equally to the overall cost.

However, this rule becomes invalid in ciphertext scenarios. In plaintext scenarios, the complexity of a convolutional layer is positively correlated with all factors—$c_i$, $c_o$, $k_w$, $k_h$, $HW$—with the equal degree of influence. However, in ciphertext scenarios, we observe that \textbf{the complexity of HE operations is influenced exclusively by specific factors rather than all of them}.
The following equations represent the complexity of in-Rot and ex-Rot, where $exRot_1$ and $exRot_2$ denote the complexity of ex-Rot in Grouped Out-Rotation MIMO and In-Rotation MIMO, and $exRot$ represents the integrated complexity:

\vspace*{-0.15in}
\begin{align}
inRot &=  \frac{c_{i}}{c_{n}}(k_{w}k_{h}-1), \label{inRot}\\
exRot_{1} &=  \frac{c_{o}}{c_{n}}(c_{n}-1), \label{exRot_out_ro}\\
exRot_{2} &=  \frac{c_{i}}{c_{n}}(c_{n}-1)k_{w}k_{h} ,\\
exRot &=  \frac{c_{n}-1}{c_{n}} \times min\langle k_{w}k_{h}c_{i} , c_{o} \rangle,
\label{decision_formula1}
\end{align}
\vspace*{-0.15in}

First, as shown in Eq.~\ref{inRot}, the complexity of in-Rot operations is solely determined by the number of input channels $c_i$ and is independent of the number of output channels $c_o$.
Next, the complexity of ex-Rot is given by Eq.~\ref{decision_formula1}, where $\min\langle a, b \rangle$ denotes the smaller of the two values, indicating that the MIMO scheme with the lower complexity is selected. Under this formulation, for layers with a kernel size of $1\times1$, the ex-Rot complexity depends on a single dimension—either $c_i$ or $c_o$—depending on which is smaller. In contrast, for layers with larger kernels, such as $3\times3$ or $5\times5$, the ex-Rot complexity becomes correlated with $c_o$, as $c_o$ is typically smaller than $9c_i$ or $25c_i$, respectively.
Overall, for any given convolutional layer, the HE Rotation complexity is governed by only one dominant factor—either $c_i$ or $c_o$. As a result, optimizing the remaining factors has negligible impact, rendering such efforts ineffective in the ciphertext setting.
A representative counterexample is DenseNet. Its high-level design aims to shift computational cost from $3\times3$ layers to $1\times1$ layers, based on the assumption that the former incurs approximately $9\times$ more FLOPs than the latter in plaintext settings. However, as previously discussed, HE rotation complexity is determined solely by either $c_i$ or $c_o$, and is independent of the kernel size $k_w\times k_h$. Consequently, this design choice offers no advantage under HE and even be detrimental, as it introduces additional HE cost in $1\times1$ layers.
Therefore, our first architectural principle is:

\noindent\textbf{$\bullet$ Focused Constraint Principle:} \textit{To constrain the factors directly impact the dominant HE complexity in order to minimize computational cost, while allowing the remaining factors to vary freely, maximizing network’s performance capacity.}
\vspace*{0.05in}

\subsubsection{\textbf{Channel Packing Aware Network Scaling}}
\label{Dimension}
During the full network evaluation for HE efficiency, we observe several distinct differences compared to the plaintext setting. First, unlike in the plaintext setting where each operation incurs a uniform computational cost, i.e., \textit{every multiply-add operation requires the same runtime} in the HE setting, \textbf{the cost of HE operations is not uniform}. As discussed earlier, in-Rot operations associated with corner weights are significantly more expensive than others. Moreover, the average costs of in-Rot and ex-Rot operations vary across the network depth. As shown in Table~\ref{AverageRotationCost}, in-Rot operations are generally more costly in shallower layers, whereas ex-Rot operations tend to dominate the computational cost in deeper layers of network.

\begin{table}[h]
\centering
\vspace*{-0.1in}
\caption{Variation in Average ex-Rot, in-Rot Cost with Model Depth (Shallow Layer$\rightarrow$Deep Layers: $c_n$=2$\rightarrow$$c_n$=512).}
\vspace*{-0.10in}
\label{AverageRotationCost}
\resizebox{70mm}{!}{
\begin{tabular}{c|ccccc}
\hline
\# $c_n$ &  2  &  \multicolumn{1}{c}{8}  &  32  &  \multicolumn{1}{c}{128}  &  \multicolumn{1}{c}{512} \\ \hline
Avg in-Rot cost(ms) &  \multicolumn{1}{l}{16.10}  &  13.13  &  \multicolumn{1}{l}{11.48}  &  9.68  &  8.82 \\ \hline
Avg ex-Rot cost(ms) &  \multicolumn{1}{l}{5.71}  &  7.50  &  \multicolumn{1}{l}{11.30}  &  14.50  &  18.93 \\ \hline
\end{tabular}}
\vspace*{-0.1in}
\end{table}

We also observe that certain layers are highly sensitive to changes in dimensionality—small increases can lead to sharp surges in HE cost. Upon further investigation, we find that the fluctuation in HE complexity also varies with network depth. These insights motivate us to identify the key contributing factors and to develop corresponding architectural design principles.

\begin{figure}[t]
\centering
\includegraphics[trim={0cm 0cm 0cm 0cm}, clip, scale=0.5]{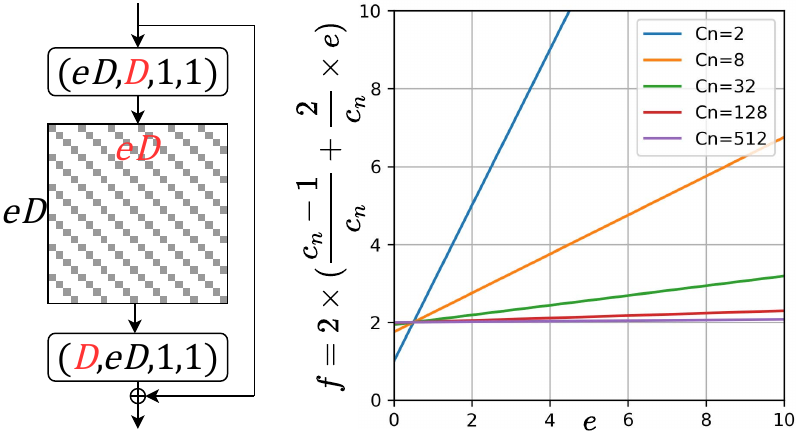}
\vspace*{-0.15in}
\caption{Dimensions of StriaBlock.}
\vspace*{-0.2in}
\label{DimensionofStriaBlock}
\end{figure}

Similar to other existing DL models constructed using fundamental components, StriaNet is architected with the proposed building block, StriaBlock. Figure~\ref{DimensionofStriaBlock} illustrates a StriaBlock on the left side. Its complexity of the HE rotation operations, which comprise both in-Rot and ex-Rot, can be expressed as:
\begin{equation}
\begin{split}
Rot_{(StriaBlock)} &=  (\frac{4}{c_{n}}\times e \times D)+(\frac{c_{n}-1 }{c_{n}}\times 2 \times D),
\end{split}
\label{exRotStriaBlock1}
\end{equation}
where $\bm{e}$ is the \textbf{scaling factor}, which represents the extension ratio of the inverse-bottleneck structure. 
Next, we simplify Eq. \ref{exRotStriaBlock1} to highlight the relationship between given dimension $D$ and the Rotation complexity of StriaBlock, resulting in:\
\begin{equation}
\begin{split}
Rot_{(StriaBlock)} &=  { 2 \times (\frac{c_{n}-1 }{c_{n}}  + \frac{2}{c_{n}}\times e )}\times D,
\end{split}
\label{exRotStriaBlock2}
\vspace*{-0.2in}
\end{equation}

\noindent The polynomial coefficient, $2 \times \left(\frac{c_{n}-1 }{c_{n}} + \frac{2}{c_{n}} \times e \right)$, represents the proportionality coefficient governing the relationship between dimension $D$ and the complexity of Rotation operations. This relationship is graphically depicted on the right side of Figure~\ref{DimensionofStriaBlock}.
In shallow layers (e.g., when $c_n=2$), even small changes in the scaling factor $e$ result in substantial increases in the proportionality coefficient. In contrast, in deeper layers (e.g., when $c_n=512$), the proportionality coefficient remains relatively stable and exhibits low sensitivity to changes in $e$.
These observations highlight that a key factor contributing to the fluctuation in HE cost across network depths is the ciphertext’s channel packing capacity, $c_n$, which varies with depth. In DL models, down-sampling progressively reduces the size of input features. Under packed HE settings, this leads to an increase in $c_n$, meaning more channels are packed into a single ciphertext. This, in turn, influences the complexity of HE operations. This insight motivates the formulation of our second architectural principle:

\noindent\textbf{$\bullet$ Channel Packing-Aware Scaling Principle:} \textit{Dynamically adjust the scaling factor $e$, the bottleneck ratio of each building block, based on the ciphertext’s channel packing capacity, $c_n$. Specifically, $e$ should be constrained in shallow layers to mitigate cost spikes and progressively increased in deeper layers to fully exploit the available network capacity.}

Following this architectural principle, we construct the network using the proposed building blocks.
Table~\ref{Architecture-L} outlines our architecture for the ImageNet dataset, featuring a scaling factor $e$ that varies from 2 to 8 in alignment with network depth. Detailed performance evaluations are presented in the evaluation section.

\vspace*{-0.2in}
\begin{table}[h]
\centering
\caption{StriaNet Architectures for ImageNet.}
\vspace*{-0.10in}
\includegraphics[trim={0cm 0cm 0cm 0cm}, clip, scale=0.64]{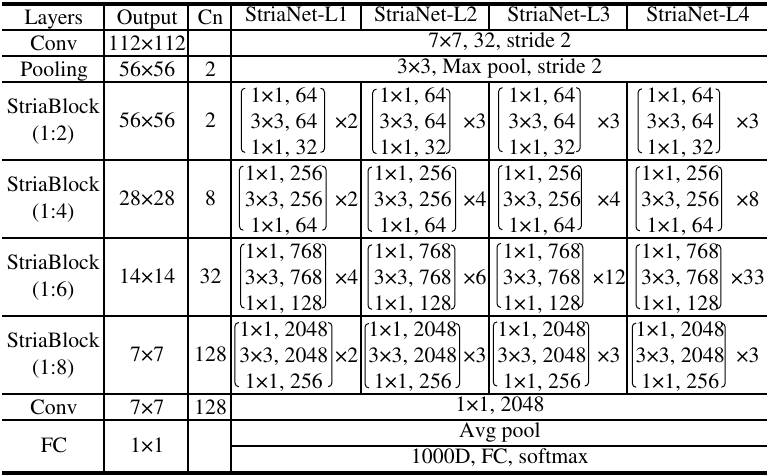}
\label{Architecture-L}
\vspace*{-0.2in}
\end{table}

\subsection{Comparing with Depthwise Convolution}
\label{Comparing_Depthwise_Convolution}

\begin{figure}[t]
\centering
\includegraphics[trim={0cm 0cm 0cm 0cm}, clip, scale=0.24]{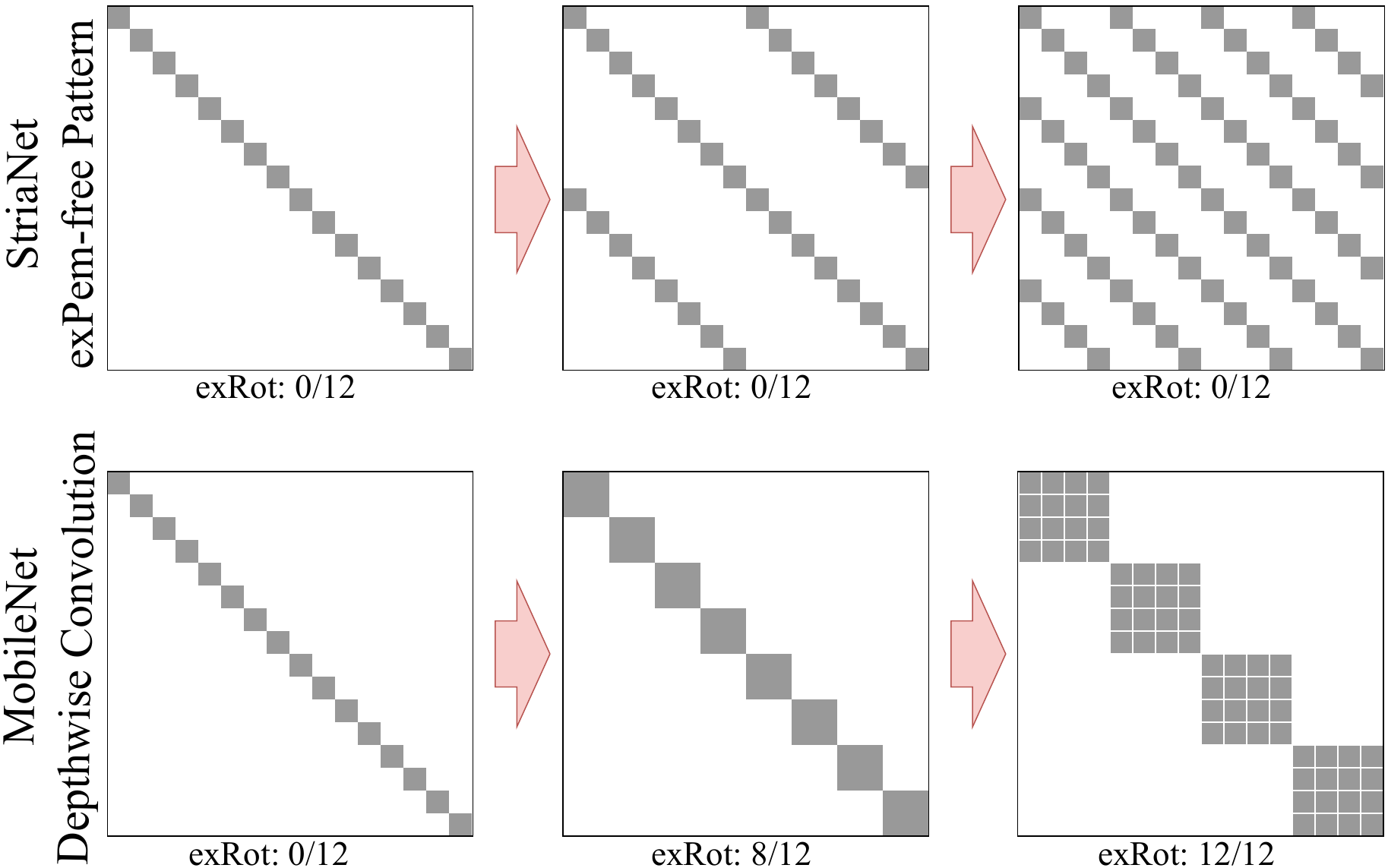}
\vspace*{-0.1in}
\caption{ExRot-Free Pattern and Depthwise Convolution.}
\vspace*{-0.2in}
\label{Stria_VS_Mobile}
\end{figure}

\textcolor{black}{The building blocks of Xception~\cite{chollet2017xception} and MobileNet~\cite{howard2017mobilenets}, when utilizing the simplest form of depthwise convolution, incidentally achieve an exRot-free configuration, corresponding to exRot-Free-pattern with $extension=1$. 
However, our exRot-Free pattern typically employs larger extensions—e.g., $extension=4$, as illustrated in Figure~\ref{Stria_VS_Mobile}. This pattern introduces additional stria patterns while consistently maintaining the exRot-free configuration. Conversely, depthwise convolution disrupts the exRot-free configuration by enhancing the main kernel diagonal, increasing kernel group sizes from $1\times1$ to $4\times4$. This disruption significantly raises the number of ex-Rot operations—from 0 ex-Rot to 8 ex-Rot, and eventually to 12 ex-Rot. In practical implementations, the extension can go as high as 32, where an extended MobileNet would require 224 ex-Rot.
This distinction arises from the fundamentally different designs of our StriaBlock and depthwise convolution. The StriaBlock featuring the exRot-Free pattern, is specifically designed for MIMO schemes in HE setting, enabling us to circumvent the resource-intensive ex-Rot operation. In contrast, depthwise convolution is primarily designed for lightweight filtering in plaintext setting, employing a single convolutional kernel per input channel to reduce the computation of feature extraction.}

\vspace*{-0.1in}
\section{Performance Evaluation}
\label{Evaluation}

All models are trained using PyTorch~\cite{NEURIPS2019_9015} without any additional enhancement techniques, such as advanced training strategies or data preprocessing methods (e.g., cleaning or augmentation), on servers with NVIDIA A100 GPUs. Detailed training configurations are provided in Appendix-\ref{Training_Setup}.
The trained StriaNet is deployed using the SEAL library~\cite{sealcrypto,SEALPython} and the Secure and Correct Inference (SCI) library from EzPC~\cite{EzPC}, evaluated on a testbed featuring an AMD 3995 CPU\footnote{Unless otherwise specified, 4 out of 128 CPU threads are utilized.}. All performance evaluations are conducted with parameters set to match the default configurations used in existing state-of-the-art frameworks~\cite{rathee2020cryptflow2,279898}, ensuring a fair comparison.
A comprehensive evaluation is carried out on three widely used datasets of varying scales: large (ImageNet~\cite{ILSVRC15}), medium (Tiny ImageNet~\cite{TinyImageNet}), and small (CIFAR-10~\cite{krizhevsky2009learning}). The analysis primarily emphasizes linear computational costs in convolutional layers, as they constitute the dominant component of end-to-end secure CNNs inference, with Rotation and Mult operations being the primary contributors.

\vspace*{-0.05in}
\subsection{ImageNet Classification}
\label{ImageNetClassification}

\begin{figure}[t]
\centering
\includegraphics[trim={0cm 0cm 0cm 0cm}, clip, scale=0.53]{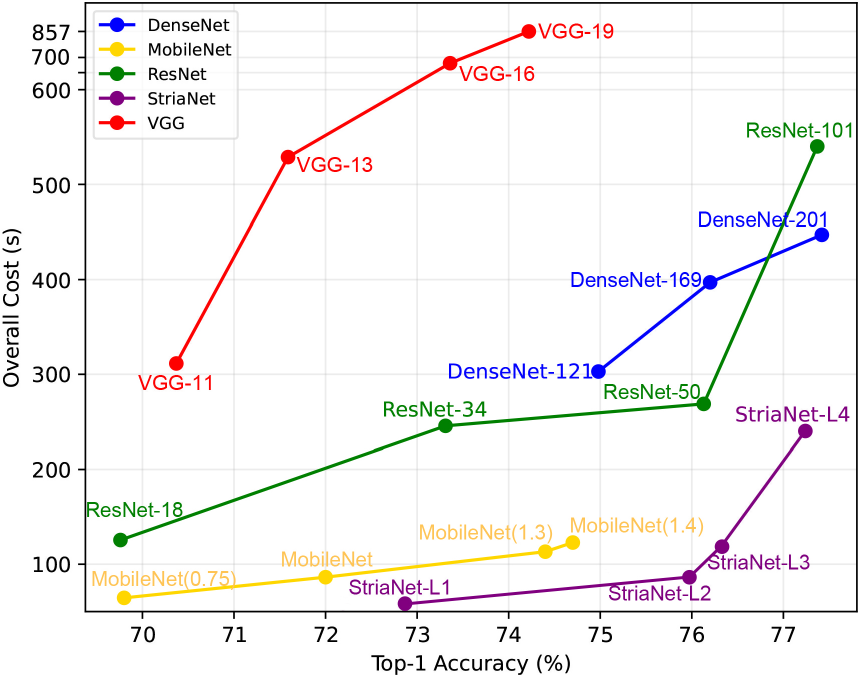}
\vspace*{-0.1in}
\caption{Performance curve of StriaNet-L Series compared to Series of VGG, ResNet, DenseNet and MobileNet. Overall costs are under LAN conditions.}
\label{eval_curve}
\vspace*{-0.2in}
\end{figure}

We evaluate the StriaNet-Large (L) Series, specifically L1-L4, on the ImageNet dataset, comparing it with classic backbone neural networks including VGG~\cite{simonyan2014very}, ResNet~\cite{he2016deep}, DenseNet~\cite{huang2017densely}, and MobileNet v2 (with varying width multipliers)\cite{sandler2018mobilenetv2, tensorflowmodelgarden2020}. 
Figure~\ref{eval_curve} offers a visual comparison, while Table~\ref{IN} summarizes the detailed performance metrics. 
At comparable accuracy levels, StriaNet achieves up to a 9.78$\times$ speedup over VGG, and 3.35$\times$ and 2.25$\times$ speedups over DenseNet and ResNet, respectively. 
Compared to the lightweight MobileNet, StriaNet demonstrates a significant performance advantage under the same cost. For instance, StriaNet-L2 outperforms MobileNet by 4\% in accuracy under equivalent resource constraints.

\begin{table*}[h]
\centering
\caption{Performance Results Comparison on ImageNet Classification.}
\vspace*{-0.1in}
\label{IN}
\resizebox{135mm}{!}{
\begin{tabular}{c|c|c|c|c|c|c|c c|c|c}
\Xhline{1.5pt}
\textbf{Models} & \textbf{Top-1 Acc} & \textbf{Params} & \textbf{\#in-Rot} & \textbf{\#ex-Rot} & \textbf{\#Mult} & \makecell{\textbf{Commu.}\\\textbf{Size}} & \multicolumn{2}{c|}{\makecell{\textbf{Overall Runtime} \\ \Xhline{0.4pt} \textbf{(LAN, WAN)}}} & \makecell{\textbf{Cost}\\\textbf{/Params}} & \makecell{\textbf{Cost}\\\textbf{/Acc}} \\ \hline
VGG-19           & 74.22\%  & 20.02M         &11,968     & 4,704          & 2,668K          &858.7MB      & 857.35s & 905.30s         & 43         & 1155             \\
VGG-16       & 73.36\%      & 14.71M         &11,584     & 3,216          & 2,447K          & 834.2MB     & 693.08s & 739.79s          & 47        & 945              \\
VGG-13            & 71.59\% & 9.4M          &9,920     & 2,144          & 1,783K          &728.0MB      & 528.81s & 570.17s          & 56         & 739            \\
VGG-11            & 70.37\% & 9.22M         &3,776     & 2,144           & 1,193K          &335.9MB      & 311.43s & 333.05s          & 34         & 443           \\ \hline
ResNet-101           & 77.37\%  & 39.63M         &\textcolor{orange}{2,848}     & \textcolor{orange}{46,548}          & 1,186K          &1107.8MB      & 540.09s & 596.33s         & \textcolor{orange}{14}     & \textcolor{orange}{698}              \\
ResNet-50       & 76.13\%      & 20.69M         &\textcolor{blue}{1,760}     & \textcolor{blue}{21,252}          & 594K          &722.9MB      & 268.76s & 305.06s          & \textcolor{blue}{13}     & \textcolor{blue}{353}                 \\
ResNet-34            & 73.31\% & 21.1M          &3,408     & 7,112          & 581K          &219.4MB      & 245.61s & 256.75s          & 12     & 335               \\
ResNet-18            & 69.76\% & 10.99M         &1,808     & 3,600           & 286K          &117.3MB      & 125.53s & 131.53s          & 11      & 180               \\ \hline
DenseNet-201         & 77.42\% & 18.72M         &\textcolor{orange}{6,400}     & \textcolor{orange}{18,636}          & 648K           &1461.8MB      & 447.22s & 505.45s          & \textcolor{orange}{24}    & \textcolor{orange}{578}                \\ 
DenseNet-169         & 76.20\% & 16.06M         &\textcolor{blue}{5,888}     & \textcolor{blue}{16,156}          & 565K           &1236.8MB      & 397.07s & 447.19s          & \textcolor{blue}{25}    & \textcolor{blue}{521}              \\ 
DenseNet-121         & 74.98\% & 9.3M         &5,504     & 10,852          & 478K           &1034.1MB      & 302.92s & 345.74s          & 33       & 404           \\ \hline
MobileNet(1.4)    & 74.70\%  & 5.07M          &4,248     & 4,546           & 108K           &655.8MB      & 122.88s & 157.91s          & 24      & 164               \\
MobileNet(1.3)    & 74.40\%  & 4.49M          &3,912     & 4,344           & 92K           &604.3MB      & 113.23s & 145.53s          & 25       & 152              \\
MobileNet(1.0)         & \textcolor{mydarkred}{72.00\%}   & 2.98M          &2,925     & 3,556          & 54K           &448.4MB      & \textcolor{mydarkred}{86.31s} & 110.06s          &29   &120                 \\
MobileNet(.75)   & 69.80\%  & 1.55M          &2,628     & 2,498         & 38K           &402.2MB      & 64.48s & 85.84s          & 42            & 92      \\ \hline
\textbf{StriaNet-L4} & \textbf{77.24\%}  & \textbf{14M}   & \textcolor{orange}{\textbf{4,768}}  & \textcolor{orange}{\textbf{10,716}} & \textbf{439K} & \textbf{1384.8MB}     & \textbf{240.14s} & \textbf{311.16s} & \textcolor{orange}{\textbf{17}}  & \textcolor{orange}{\textbf{311}}\\
\textbf{StriaNet-L3} & \textbf{76.33\%}  & \textbf{7.65M} & \textcolor{blue}{\textbf{2,240}}  & \textcolor{blue}{\textbf{5,060}} & \textbf{213K}  & \textbf{677.2MB}     & \textbf{118.53s} & \textbf{154.06s} & \textcolor{blue}{\textbf{15}}  & \textcolor{blue}{\textbf{155}}\\
\textbf{StriaNet-L2} & \textcolor{mydarkred}{\textbf{76.02\%}}  & \textbf{5.92M} & \textbf{1,664}  & \textbf{3,572} & \textbf{159K}  & \textbf{517.9MB}     & \textcolor{mydarkred}{\textbf{87.68s}} & \textbf{115.19s} & \textbf{15}   & \textbf{115}\\
\textbf{StriaNet-L1} & \textbf{72.79\%}  & \textbf{3.44M} & \textbf{1,024}  & \textbf{2,312}  & \textbf{106K}  & \textbf{333.1MB}     & \textbf{58.64s} & \textbf{76.84s} & \textbf{15}  & \textbf{80}\\ \Xhline{1.5pt}
\end{tabular}}
\\\footnotesize{*The network bandwidth: LAN 3Gbps WAN: 400Mbps.}
\vspace*{-0.2in}
\end{table*}

In addition to overall performance, our architecture, designed with a strong focus on Rotation optimizations, achieves significantly lower Rotation complexity compared to networks originally designed for plaintext scenarios. As shown in Table~\ref{IN}, several groups of Rotation counts are highlighted in different colors to illustrate these comparisons.

Table~\ref{IN} also presents two specific performance metrics: the HE cost per unit parameter (Cost/Params) and the HE cost per unit performance (Cost/Acc). These metrics provide deeper insights into each models' efficiency and cost effectiveness under HE scenarios. For fairness, comparisons are also conducted at matched accuracy levels, indicated by different color annotations.
For the \textbf{Cost/Params} metric, VGG unsurprisingly offers no advantage. DenseNet and MobileNet exhibit nearly 200\% the inefficiency compared to StriaNet in this regard. Notably, ResNet achieves a similar level of per-parameter efficiency to StriaNet. However, due to its significantly larger parameter count, approximately 2.8$\times$ more, it incurs a substantially higher overall HE cost.
It is also noteworthy that although MobileNet exhibits the lowest total HE cost among the four plaintext networks, its relatively high Cost/Params value suggests limited scalability. This implies that as the model size and parameter count grow, the HE cost of MobileNet will rise rapidly, potentially surpassing even that of ResNet and further widening the efficiency gap with StriaNet.
For the \textbf{Cost/Acc} metric, StriaNet demonstrates a clear comparative advantage over all competing plaintext networks. Its HE cost per unit of performance can be as low as 29.75\% of the baseline, underscoring the exceptional efficiency of networks specifically designed for HE settings compared to plaintext-originated architectures.

\begin{table}[h]
\centering
\vspace*{-0.15 in}
\caption{Building Block Complexity Comparison}
\vspace*{-0.1in}
\label{layerwise_buildingblock}
\resizebox{70mm}{!}{
\begin{tabular}{c|c|c|c|c|c}
\Xhline{1.5pt}
\multicolumn{1}{c|}{Depth}            &\multicolumn{1}{c|}{Input}             & Metric           & ResNet                      & MobileNet                  & \textbf{StriaNet}                   \\ \hline
\multicolumn{1}{c|}{{\color[HTML]{FF0000} }}       &\multicolumn{1}{c|}{{\color[HTML]{FF0000} }}           & \#in-Rot       & 256                         & 576                        & 128                        \\
\multicolumn{1}{c|}{{\color[HTML]{FF0000} }}       &\multicolumn{1}{c|}{{\color[HTML]{FF0000} }}              & \#ex-Rot       & 96                          & 24                         & 32                         \\ 
\multicolumn{1}{c|}{\multirow{-2.8}{*}{{\rotatebox{90}{\textit{Deep Blocks  $\leftarrow$  Shallow Blocks}}}}}  &\multicolumn{1}{c|}{\multirow{-2.7}{*}{{\rotatebox{90}{\textbf{56$\times$56}}}}}       & \textbf{\# Rotation} & {\color[HTML]{000000} 352}  & {\color[HTML]{000000} 600} & {\color[HTML]{000000} \textbf{160}} \\ \cline{3-6}    
\cline{2-6} 
\multicolumn{1}{c|}{{\color[HTML]{FF0000} }}       & \multicolumn{1}{c|}{{\color[HTML]{FF0000} }}                         & \#in-Rot       & 128                         & 192                        & 128                        \\
\multicolumn{1}{c|}{{\color[HTML]{FF0000} }}       & \multicolumn{1}{c|}{{\color[HTML]{FF0000} }}                         & \#ex-Rot       & 336                          & 56                         & 112                         \\ 
\multicolumn{1}{c|}{{\color[HTML]{FF0000} }}  & \multicolumn{1}{c|}{\multirow{-2.7}{*}{{\rotatebox{90}{\textbf{28$\times$28}}}}}     & \textbf{\# Rotation} & {\color[HTML]{000000} 464}  & {\color[HTML]{000000} 248} & {\color[HTML]{000000} \textbf{240}} \\ \cline{3-6} 
\cline{2-6} 
\multicolumn{1}{c|}{{\color[HTML]{FF0000} }}       & \multicolumn{1}{c|}{{\color[HTML]{FF0000} }}                         & \#in-Rot       & 64                         & 144                        & 96                        \\
\multicolumn{1}{c|}{{\color[HTML]{FF0000} }}       & \multicolumn{1}{c|}{{\color[HTML]{FF0000} }}                         & \#ex-Rot       & 744                          & 186                         & 248                         \\ 
\multicolumn{1}{c|}{{\color[HTML]{FF0000} }}   & \multicolumn{1}{c|}{\multirow{-2.7}{*}{{\rotatebox{90}{\textbf{14$\times$14}}}}}                         & \textbf{\# Rotation} & {\color[HTML]{000000} 808}  & {\color[HTML]{000000} 330} & {\color[HTML]{000000} \textbf{344}} \\ \cline{3-6} 
\cline{2-6} 
\multicolumn{1}{c|}{{\color[HTML]{FF0000} }}       & \multicolumn{1}{c|}{{\color[HTML]{FF0000} }}                         & \#in-Rot       & 32                         & 64                        & 64                        \\
\multicolumn{1}{c|}{{\color[HTML]{FF0000} }}       & \multicolumn{1}{c|}{{\color[HTML]{FF0000} }}                        & \#ex-Rot       & 1524                          & 508                         & 508                         \\ 
\multicolumn{1}{c|}{{\color[HTML]{FF0000} }}   & \multicolumn{1}{c|}{\multirow{-2.7}{*}{{\rotatebox{90}{\textbf{7$\times$7}}}}}            & \textbf{\# Rotation} & {\color[HTML]{000000} 1556}  & {\color[HTML]{000000} 572} & {\color[HTML]{000000} \textbf{572}} \\ \cline{3-6} 
\Xhline{1.5pt}
\end{tabular}}
\vspace*{-0.15in}
\end{table}

\vspace*{0.05in}\noindent\textbf{Building Block Complexity Benchmark.} 
As ResNet, MobileNet, and StriaNet all adopt building blocks as their fundamental architectural units, we conduct a benchmark evaluation of their respective blocks. The results, presented in Table~\ref{layerwise_buildingblock}, compare their HE Rotation complexity across different depths, ranging from shallow network with large input ($56\times56$) to deep network with smaller input sizes ($7\times7$). 
For both ResNet and StriaNet, the Rotation complexity of their building blocks increases with network depth. However, due to optimized design such as the exRot-Free pattern, StriaNet consistently exhibits lower Rotation complexity.
In contrast, MobileNet's building blocks show noticeable spikes of Rotation complexity in shallow layers, primarily due to the high number of in-Rot. StriaNet, on the other hand, maintains stable Rotation complexity throughout the network. This stability is attributed to its dynamic scaling factor, which adjusts the bottleneck ratio of each building block with network depths, combined with in-Rot optimization provided by the Cross Kernel design.

\begin{table}[h]
\centering
\vspace*{-0.15in}
\caption{Ablation Study}
\vspace*{-0.1in}
\label{Ablation}
\resizebox{65mm}{!}{
\begin{tabular}{c|ccc|c}
\Xhline{1.2pt}
                             & CPAS & ExRot-free & Cross Kernel & times(s)  \\ \hline
\multirow{4}{*}{StriaNet-L3} & -    & -          & -           & 1797.91 \\
                             & $\checkmark$    & -          & -           & 971.20 (46\%$\downarrow$)  \\
                             & $\checkmark$    & $\checkmark$          & -           & 169.37 (83\%$\downarrow$)  \\
                             & $\checkmark$    & $\checkmark$          & $\checkmark$           & \textbf{118.53} (30\%$\downarrow$)  \\ \Xhline{1.2pt}
\end{tabular}}
\\\footnotesize{*The ``$\downarrow$'' denotes the relative reduction compared to the row above.}
\vspace*{-0.1in}
\end{table}

\vspace*{0.05in}\noindent\textbf{Ablation Study.} 
Although architecture functions as a unified system with jointly contributing components, the ablation study highlights the individual contributions of each design element. As shown in Table~\ref{Ablation}, the ExRot-free Pattern contributes significantly to improvements. Moreover, Channel Packing-Aware Scaling (CPAS) enhances the network by optimizing dimensional configuration, resulting in twofold enhancement. Lastly, Cross Kernel further boosts model efficiency by specifically targeting in-Rot optimization.

\begin{table}[h]
\centering
\vspace*{-0.15in}
\caption{Peak Memory Usage for Ciphertext Storage}
\vspace*{-0.1in}
\label{Peak_Memory}
\resizebox{60mm}{!}{
\begin{tabular}{c|c|c}
\Xhline{1.5pt}
Model                   & $HW, c_o, c_i, k, s$     & Memory Overhead  \\ \hline
ResNet                        & $56^2,64,64,3,1$  & 324MB                     \\
DenseNet                         & $56^2,32,128,3,1$  & 648MB                  \\
MobileNet                         & $56^2,144,144,3,1$ & 729MB                     \\
\textbf{StriaNet}                         & $56^2,64,64,3,1$   & \textbf{180MB}                  \\ 
\Xhline{1.5pt}
\end{tabular}}
\vspace*{-0.1in}
\end{table}


\noindent\textbf{Memory Overhead.} 
Memory overhead in secure inference is measured by peak memory usage, related to the maximum number of ciphertexts stored simultaneously. Table~\ref{Peak_Memory} presents memory overhead for the implementations of ResNet, MobileNet, DenseNet, and StriaNet with the ImageNet dataset. The second column details where the peak memory occurs, along with the corresponding configurations.
Compared to other architectures, StriaNet requires only 180MB of memory, demonstrating a significantly reduced memory footprint.

\vspace*{-0.15in}
\begin{table}[h]
\centering
\caption{Results on Tiny ImageNet, CIFAR-10 Classification.}
\vspace*{-0.1in}
\label{ACC_Tiny_ImageNet_Cifar}
\resizebox{80mm}{!}{
\begin{tabular}{cc|c|c|c|c}
\Xhline{1.5pt}
\multicolumn{2}{c|}{\textbf{Models}}                                        & \textbf{Top-1 Acc}    & \textbf{Rotation (s)} & \textbf{Mult (s)} & \textbf{Overall (s)} \\ \hline
\multicolumn{1}{c|}{\multirow{7}{*}{\rotatebox{90}{Tiny ImageNet}}} & ResNet-50        & 65.50\%  & 150.69           & 140.00       & 254.69      \\
\multicolumn{1}{c|}{}                               & \textit{ResNet-50-Pruned}   & 65.54\% & 70.29           & 36.13     &106.43       \\
\multicolumn{1}{c|}{}                               & VGG-16   & 65.00\% & 77.89           & 254.31        & 323.20   \\
\multicolumn{1}{c|}{}                               & \textit{VGG-16-Pruned}   & 64.94\% & 41.36          & 56.01      & 97.37     \\
\multicolumn{1}{c|}{}                               & MobileNetV2      & 67.37\% & 107.55           & 21.91      & 129.46    \\
\multicolumn{1}{c|}{}                               & \textbf{StriaNet-M2}      & \textbf{68.57\%}          & \textbf{51.91}        & \textbf{31.67}  & \textbf{83.58} \\ 
\multicolumn{1}{c|}{}                               & \textbf{StriaNet-M1}      & \textbf{67.49\%}          & \textbf{34.63}        & \textbf{19.13}   & \textbf{53.76}\\ 
\Xhline{1.0pt}
\multicolumn{1}{c|}{\multirow{6}{*}{\rotatebox{90}{CIFAR-10}}}      & ResNet-50       & 93.50\%               & 172.35          & 27.79       &  200.14    \\
\multicolumn{1}{c|}{}                               & VGG-16                         & 94.50\%            & 58.37           & 61.34          & 119.72\\
\multicolumn{1}{c|}{}                               & \textit{VGG-16-Pruned}        & 94.41\%              & 25.29           & 16.40           & 41.69\\
\multicolumn{1}{c|}{}                               & ResNet-56-CIFAR             & 93.03\%              & 47.38           & 13.57           & 60.95\\
\multicolumn{1}{c|}{}                               & \textbf{StriaNet-S2}     & \textbf{95.14\%}   & \textbf{24.96}            & \textbf{7.82}       & 32.78    \\ 
\multicolumn{1}{c|}{}                               & \textbf{StriaNet-S1}     & \textbf{94.56\%}   & \textbf{13.43}            & \textbf{8.24}       & 21.67    \\ 
\Xhline{1.5pt}
\end{tabular}}
\vspace*{-0.2in}
\end{table}

\subsection{Tiny ImageNet and CIFAR-10 Classification}
\label{TinyClassification}


\noindent We also evaluate StriaNet's performance on the Tiny ImageNet and CIFAR-10 datasets, as presented in Table~\ref{ACC_Tiny_ImageNet_Cifar}. On both benchmarks, StriaNet demonstrates exceptional cost efficiency, achieving speedups of $6.01\times$ and $9.24\times$, respectively. Notably, compared to pruned ResNet-50~\cite{cai2024mosaic}, StriaNet is \textbf{200\%} faster with improving accuracy by $2\%$. This underscores the effectiveness of architectures tailored for HE settings over those adapted from plaintext models, establishing StriaNet as a new benchmark for cost-efficient HE inference. For architectural details of the StriaNet-M and StriaNet-S series and more performance details, please refer to Appendix-\ref{Architecture_tiny_cifar}.

\subsection{Exploration of StriaNet with Cheetah}
\label{Cheetah}

An important design focus of our work is minimizing HE Rotations. Cheetah~\cite{juvekar2018gazelle} adopts a specialized encoding and polynomial multiplication scheme to perform convolutions directly, thereby implicitly achieving element alignment without explicit Rotation. However, this prevents efficient arrangement and alignment of elements within the output ciphertext, resulting in low ciphertext utilization and substantial communication overhead~\cite{hao2022iron, gupta2022llama, xu2023falcon, pang2023bolt}. 

\vspace*{-0.1in}
\begin{figure}[h]
\centering
\includegraphics[trim={0cm 0cm 0cm 0cm}, clip, scale=0.38]{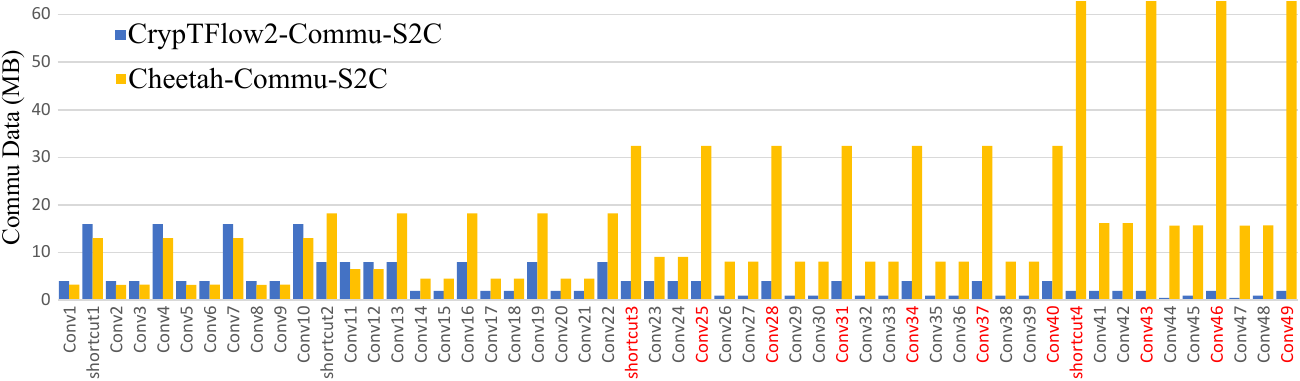}
\vspace*{-0.1in}
\caption{Layer-wise Communication. ResNet-50.}
\label{Layer-wise_Communication}
\vspace*{-0.1in}
\end{figure}

Figure~\ref{Layer-wise_Communication} presents the layer-wise communication data transmitted from the server to the client. It is evident that, as the model depth increases, Cheetah incurs significantly higher communication overhead compared to CrypTFlow2—reaching up to 31.4$\times$ more for certain layers (highlighted in red). 
This overhead can cause substantial delays in real-world applications, particularly in scenarios with limited network bandwidth, such as WAN or wireless networks. However, replacing Cheetah with original CrypTFlow2 introduces increased computational costs due to the heavy use of Rotation operations, as shown in Table~\ref{benchmarkvsCheetah}. Our Rotation-Free Pattern provides a promising solution to this trade-off, offering an architecture that minimizes HE Rotations without incurring additional communication overhead.

\vspace*{-0.1in}
\begin{table}[h]
\centering
\caption{Microbenchmarks.}
\vspace*{-0.1in}
\label{benchmarkvsCheetah}
\resizebox{70mm}{!}{
\begin{tabular}{cc|cc}
\Xhline{1.5pt}
Method                   & $HW, c_o, c_i, k, s$     & \begin{tabular}[c]{@{}c@{}}Comp. Time\\ (Sec)\end{tabular} & \begin{tabular}[c]{@{}c@{}}S2C Commu.\\ (MB)\end{tabular} \\ \hline
\multirow{4}{*}{Cheetah} & $28^2,1024,512,1,2$  & 2.92           & 32.4          \\
                         & $14^2,1024,256,1,1$  & 1.58           & 32.4          \\
                         & $14^2,2048,1024,1,2$ & 3.14           & 62.8          \\
                         & $7^2,2048,512,1,1$   & 1.86           & 62.8          \\ \hline
\multirow{4}{*}{CrypTFlow2}     & $28^2,1024,512,1,2$  & 6.25           & 4.0           \\
                         & $14^2,1024,256,1,1$  & 3.45           & 4.0           \\
                         & $14^2,2048,1024,1,2$ & 7.34           & 2.0           \\
                         & $7^2,2048,512,1,1$   & 4.54           & 2.0           \\ \hline
\multirow{4}{*}{\textbf{Ours}}    & $28^2,1024,512,1,2$  & 3.12           & 4.0           \\
                         & $14^2,1024,256,1,1$  & 1.51           & 4.0           \\
                         & $14^2,2048,1024,1,2$ & 3.21           & 2.0           \\
                         & $7^2,2048,512,1,1$   & 1.77           & 2.0           \\ \hline
\Xhline{1.5pt}
\end{tabular}}
\vspace*{-0.1in}
\end{table}

\begin{table}[h]
\centering
\caption{Performance Comparison.}
\vspace*{-0.1in}
\label{vsCheetah}
\resizebox{84mm}{!}{
\begin{tabular}{cccccc}
\Xhline{1.5pt}
\multirow{2}{*}{\begin{tabular}[c]{@{}c@{}}Model\\ Data\end{tabular}}      & \multirow{2}{*}{\begin{tabular}[c]{@{}c@{}}Server\\ CPU Thrds\end{tabular}} & \multicolumn{4}{c}{Comp \& Commu Runtime (Sec): \textbf{Ours}/Vanilla Cheetah}                     \\ \cline{3-6} 
                                                                           &                                                                             & WAN                & 5G                 & 4G/LTE             & IoT                \\ \hline
\multirow{6}{*}{\begin{tabular}[c]{@{}c@{}}RsNt50\\ cfr10\end{tabular}}    & \multirow{2}{*}{$4\times$}                                                         & 72.67/81.46        & 83.29/100.46       & 104.53/138.47      & 231.97/366.49      \\
                                                                           &                                                                             & \textbf{(89.21\%)} & \textbf{(82.91\%)}  & \textbf{(75.49\%)} & \textbf{(63.30\%)} \\ \cline{2-6} 
                                                                           & \multirow{2}{*}{$8\times$}                                                         & 42.12/50.71        & 52.74/69.71        & 73.98/107.71       & 201.42/335.74      \\
                                                                           &                                                                             & \textbf{(83.06\%)} & \textbf{(75.66\%)} & \textbf{(68.68\%)} & \textbf{(59.99\%)} \\ \cline{2-6} 
                                                                           & \multirow{2}{*}{$16\times$}                                                        & 26.61/35.10        & 37.23/54.10        & 58.47/92.11        & 185.91/320.13      \\
                                                                           &                                                                             & \textbf{(75.82\%)} & \textbf{(68.82\%)} & \textbf{(63.48\%)} & \textbf{(58.07\%)} \\ \hline
\multirow{6}{*}{\begin{tabular}[c]{@{}c@{}}RsNt50\\ tiny IN\end{tabular}}  & \multirow{2}{*}{$4\times$}                                                         & 76.46/85.18        & 88.91/106.01       & 113.81/147.68      & 263.21/397.66      \\
                                                                           &                                                                             & \textbf{(89.76\%)} & \textbf{(83.87\%)} & \textbf{(77.07\%)} & \textbf{(66.19\%)} \\ \cline{2-6} 
                                                                           & \multirow{2}{*}{$8\times$}                                                         & 44.94/53.50        & 57.39/74.33        & 82.29/115.99       & 231.69/365.98      \\
                                                                           &                                                                             & \textbf{(84.01\%)} & \textbf{(77.21\%)} & \textbf{(70.95\%)} & \textbf{(63.31\%)} \\ \cline{2-6} 
                                                                           & \multirow{2}{*}{$16\times$}                                                        & 28.95/37.42        & 41.40/58.25        & 66.30/99.91        & 215.70/349.90      \\
                                                                           &                                                                             & \textbf{(77.36\%)} & \textbf{(71.07\%)} & \textbf{(66.35\%)} & \textbf{(61.65\%)} \\ \hline
\multirow{6}{*}{\begin{tabular}[c]{@{}c@{}}RsNt101\\ tiny IN\end{tabular}} & \multirow{2}{*}{$4\times$}                                                         & 143.97/162.74      & 165.69/201.76      & 209.15/279.80      & 469.86/748.07      \\
                                                                           &                                                                             & \textbf{(88.47\%)} & \textbf{(82.12\%)} & \textbf{(74.75\%)} & \textbf{(62.81\%)} \\ \cline{2-6} 
                                                                           & \multirow{2}{*}{$8\times$}                                                         & 83.78/101.82       & 105.50/140.84      & 148.96/218.89      & 409.67/687.16      \\
                                                                           &                                                                             & \textbf{(82.28\%)} & \textbf{(74.91\%)} & \textbf{(68.05\%)} & \textbf{(59.62\%)} \\ \cline{2-6} 
                                                                           & \multirow{2}{*}{$16\times$}                                                        & 53.23/70.91        & 74.96/109.93       & 118.41/187.98      & 379.12/656.25      \\
                                                                           &                                                                             & \textbf{(75.07\%)} & \textbf{(68.19\%)} & \textbf{(62.99\%)} & \textbf{(57.77\%)} \\
\Xhline{1.5pt}
\end{tabular}}
\footnotesize{*The CPU contains a total of 128$\times$ threads. The speed of each network is: WAN: 400Mbps, 5G: 200Mbps, 4G/LTE: 100Mbps, IoT: 25Mbps~\cite{speedtest}.}
\vspace*{-0.15in}
\end{table}

The end-to-end evaluations are summarized in Table~\ref{vsCheetah}, where we replace Cheetah with our Rotation-Free Pattern for layers exhibiting high communication overhead, evaluated under varying computational capacities and real-world network conditions~\cite{speedtest}. The results show that our design effectively alleviates Cheetah’s communication bottleneck in practical deployment scenarios, achieving an overall \textbf{1.73}\bm{$\times$} speedup in end-to-end inference.

Note that work~\cite{lu2023bumblebee} mitigates Cheetah’s communication issue through an additional post-processing ``interleave" step, which fundamentally involves rotations over polynomial coefficients~\cite{hou2023ciphergpt}. In contrast, our approach fully leverages the optimization potential of standard HE protocol by employing the Rotation-Free Pattern to achieve Rotation optimization inherently. This results in a more concise and direct solution. Furthermore, our design does not rely on any specialized protocol, offering improved compatibility and generality.

\section{Conclusion and Future Work}
\label{conclusion}

This work addresses a critical gap in current privacy-preserving MLaaS research—the lack of neural network architectures explicitly designed for HE settings. We argue that significant efficiency gains can be achieved by designing networks specifically tailored to the unique computational constraints of HE, rather than adapting existing plaintext models.
We begin by introducing StriaBlock, a novel building block targeting the most computationally expensive HE operation—rotation. StriaBlock integrates ExRot-Free Convolution with a novel Cross Kernel, entirely eliminating the need for external Rotations and reducing internal Rotation usage to just 19\% of that required by plaintext models.
Next, we propose the architectural principle, includes the Focused Constraint Principle, which limits cost-sensitive factors while preserving flexibility in others, and the Channel Packing-Aware Scaling Principle, which dynamically adapts bottleneck ratios based on ciphertext channel capacity that varies with network depth. These strategies efficiently control the local and overall HE cost, enabling a balanced architecture for HE settings.
The resulting network, StriaNet, is comprehensively evaluated and achieves comparable accuracy while delivering speedups of \textbf{9.78}\bm{$\times$}, \textbf{6.01}\bm{$\times$}, and \textbf{9.24}\bm{$\times$} on ImageNet, Tiny ImageNet, and CIFAR-10, respectively.
For future work, we aim to further enhance StriaNet by exploring the incorporation of Neural Architecture Search techniques.

\section*{Acknowledgment}
The authors would like to express thanks of gratitude to the lab members for helping and collaborating on this work and anonymous reviewers for the constructive comments. The work of H. Wu was supported in part by the National Science Foundation under Grants SaTC-2439013, CNS-2413009, DGE-2336109, OAC-2320999, IIS-2236578, and CNS-2120279.



\bibliographystyle{IEEEtran}
\bibliography{sample-base}

\appendix

\subsection{Input Rotation MIMO}
\label{moreMIMO}

\begin{figure}[b]
\centering
\includegraphics[trim={0cm 0cm 0cm 0cm}, clip, scale=0.47]{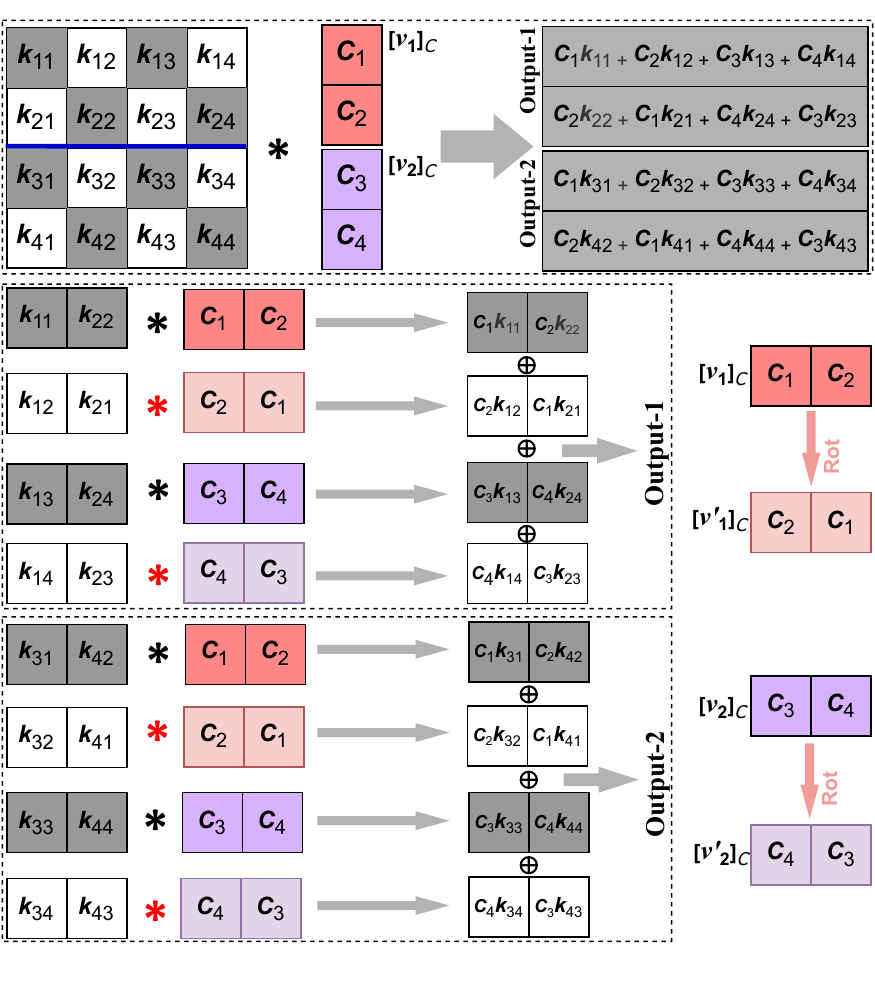}
\vspace*{-0.15in}
\caption{In-Rot MIMO ($c_n =2$).}
\label{Input-Ro}
\vspace*{-0.15in}
\end{figure}

Figure~\ref{Input-Ro} illustrates the \textbf{Input Rotation (In-Rot) MIMO} scheme, where the input of convolution, instead of the output of convolution, is rotated.
For example, the inputs of the convolution, denoted as ciphertext $[\bm{v_1}]_{\mathcal{C}}$ and $[\bm{v_2}]_{\mathcal{C}}$, are rotated into $[\bm{v'_1}]_{\mathcal{C}}$ and $[\bm{v'_2}]_{\mathcal{C}}$. Subsequently, the rotated inputs are convolved with the kernel group $\{\bm{k}_{12}, \bm{k}_{21}\}$ and $\{\bm{k}_{14}, \bm{k}_{23}\}$ to directly obtain the intermediate ciphertext $[\bm{c}_2\bm{k}_{12},\bm{c}_1\bm{k}_{21}]_{\mathcal{C}}$ and $[\bm{c}_4\bm{k}_{14},\bm{c}_3\bm{k}_{23}]_{\mathcal{C}}$.
In this way, combined with additional two intermediate ciphertext: $[\bm{c}_1\bm{k}_{11},\bm{c}_2\bm{k}_{22}]_{\mathcal{C}}$, $[\bm{c}_3\bm{k}_{13},\bm{c}_4\bm{k}_{24}]_{\mathcal{C}}$, the first output ciphertext, Output-1, is produced by summing up all four intermediate ciphertext using the Add operation. The process for obtaining the second output ciphertext, Output-2, follows a similar procedure.
It's important to note that one Rotation operation is involved in converting the input ciphertext $[\bm{v_1}]_{\mathcal{C}}$ into $[\bm{v'_1}]_{\mathcal{C}}$. Following this, an additional $(k_wk_h-1)$ in-Rot operations are performed over $[\bm{v'_1}]_{\mathcal{C}}$ to obtain the rotated ciphertexts necessary for the convolution (represented by the red \textcolor{red}{$\ast$} convolution operator) between $[\bm{v'_1}]_{\mathcal{C}}$ and the kernel group $\{\bm{k}_{12}, \bm{k}_{21}\}$, as well as all other kernel groups to be convolved with $[\bm{v'_1}]_{\mathcal{C}}$.
Obviously, all expensive Rotation operations are related to the kernel group marked in white. The proposed exRot-Free Pattern, which involves only the kernel group marked in gray, does not require any Rotation operations for calculation.

\subsection{Block-Wise Performance Breakdown}
\label{Block-Wise_Breakdown}


\begin{figure}[t]
\centering
\includegraphics[trim={0cm 0cm 0cm 0cm}, clip, scale=0.76]{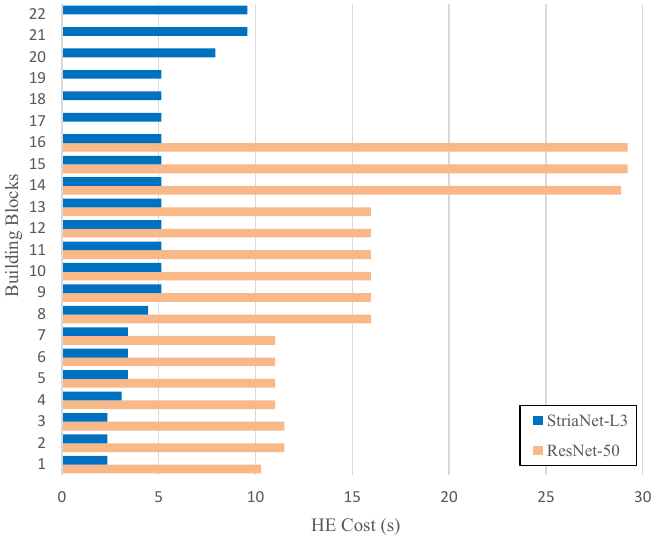}
\vspace*{-0.15in}
\caption{Block-Wise Performance Breakdown: Comparison between StriaNet-L3 and ResNet.}
\label{Block-Wise_res}
\end{figure}


\begin{figure}[t]
\centering
\includegraphics[trim={0cm 0cm 0cm 0cm}, clip, scale=0.78]{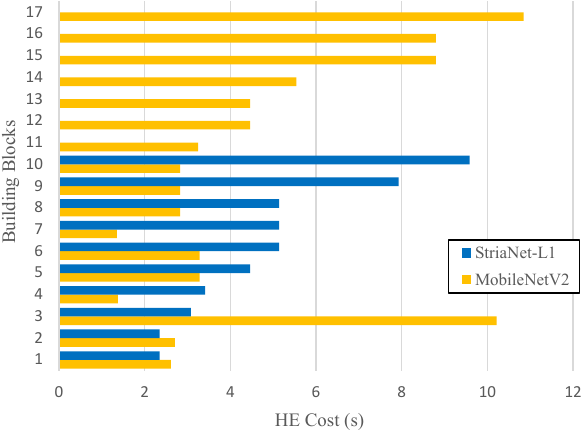}
\vspace*{-0.15in}
\caption{Block-Wise Performance Breakdown: Comparison between StriaNet-L1 and MobileNet.}
\label{Block-Wise_mobile}
\vspace*{-0.15in}
\end{figure}

Comparing StriaNet-L3 with ResNet-50 (see Figure~\ref{Block-Wise_res}), StriaNet consistently maintains significantly lower HE cost across all layers, particularly in deep layers, resulting in a 44.1\% of ResNet-50's cost.
Contrasting with the lightweight MobileNet (see Figure\ref{Block-Wise_mobile}), StriaNet-L1 exhibits a notable advantage over MobileNet in the HE cost of shallow layers due to superior dimension design aligned with unique HE characteristics. Additionally, despite higher accuracy, StriaNet requires fewer building blocks than MobileNet, i.e., 10 v.s. 17, resulting in an overall cost approximately 2/3 of MobileNet. This efficiency stems from HE-Efficient designs, allowing StriaNet to maintain a comparable Rotation complexity and HE cost as MobileNet while accommodating more effective parameters for enhanced capacity.

\subsection{StriaNet on Tiny ImageNet and CIFAR-10 Classification}
\label{Architecture_tiny_cifar}

We extended our investigations using the Tiny ImageNet and CIFAR-10 datasets to evaluate StriaNet's performance with moderate to small-scale datasets. The network architectures are outlined in Table~\ref{Architecture-M} and Table~\ref{Architecture-S}. Our focus lies on understanding StriaNet's behavior across diverse dataset scales rather than aggressively pursuing state-of-the-art results. Nevertheless, these additional studies provide valuable insights into the adaptability and efficacy of StriaNet across varying dataset scales.

\begin{table}[t]
\vspace*{-0.1in}
\centering
\caption{StriaNet Architectures for Tiny ImageNet.}
\vspace*{-0.12in}
\resizebox{65mm}{!}{
\begin{tabular}{ c }
\includegraphics[width=\linewidth]{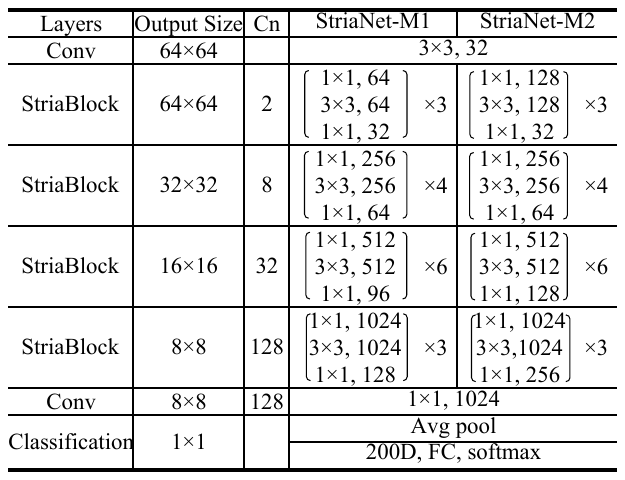}
\end{tabular}}
\label{Architecture-M}
\vspace*{-0.15in}
\end{table}


\begin{table}[t]
\centering
\caption{StriaNet Architectures for CIFAR-10.}
\vspace*{-0.12in}
\resizebox{65mm}{!}{
\begin{tabular}{ c }
\includegraphics[width=\linewidth]{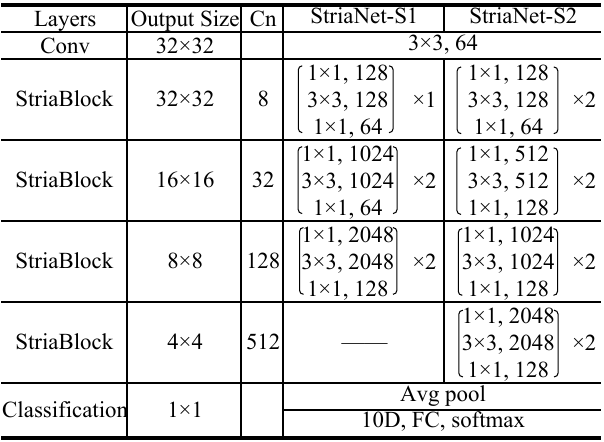}
\end{tabular}}
\label{Architecture-S}
\vspace*{-0.1in}
\end{table}

\begin{figure}[b]
\centering
\includegraphics[trim={0cm 0cm 0cm 0cm}, clip, scale=0.12]{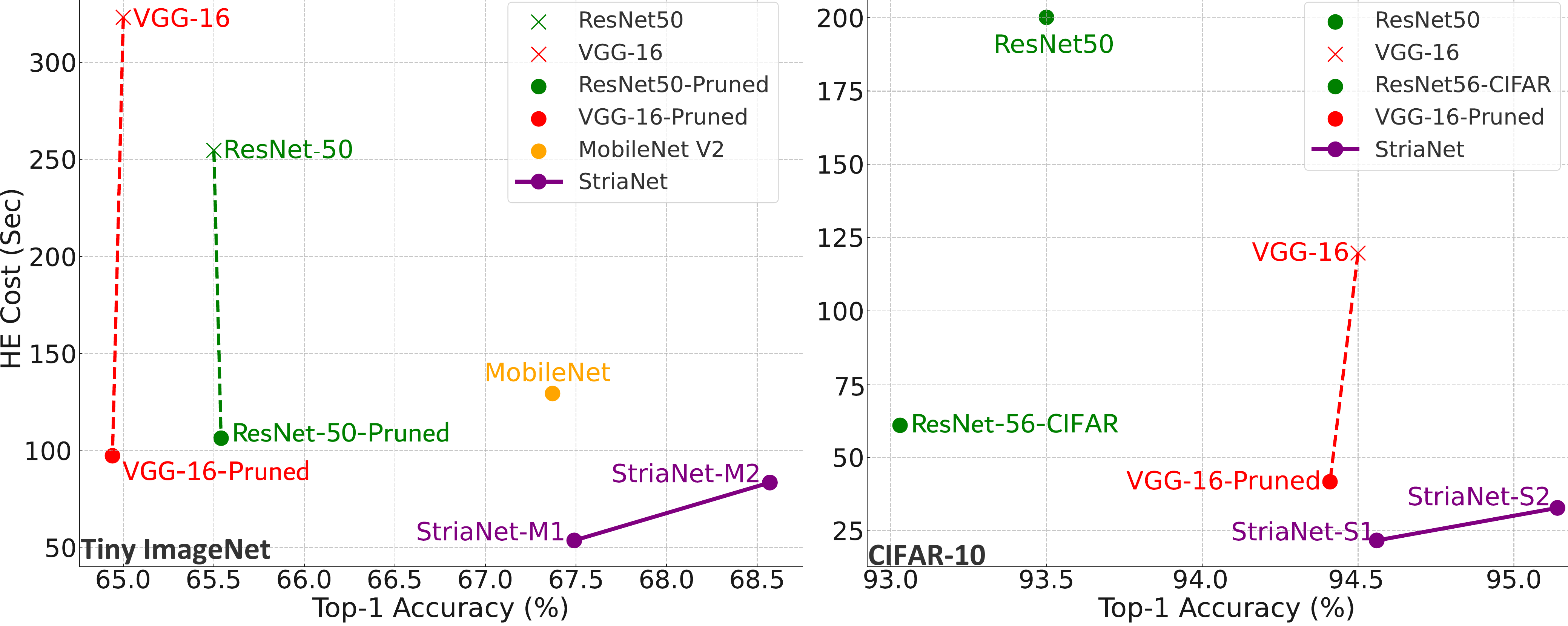}
\caption{Performance curve of StriaNet-M and StriaNet-S Series on Tiny ImageNet and CIFAR-10 Classification.}
\label{Flops_HE_tiny_cifar}
\vspace*{-0.1in}
\end{figure}

For the Tiny ImageNet dataset, shown in Table~\ref{ACC_Tiny_ImageNet_Cifar} and Figure~\ref{Flops_HE_tiny_cifar}, StriaNet is compared with the original ResNet-50, a pruned ResNet-50 using the state-of-the-art HE-friendly pruning method~\cite{cai2022hunter}, VGG-16, pruned VGG-16 and MobileNetV2. First, we introduce StriaNet-M2, achieved by simplifying the StriaNet-L Series. StriaNet-M2 demonstrates remarkable results, achieving the $3.05\times$, $1.27\times$, $3.87\times$, $1.16\times$ and $1.55\times$ speedup compared to ResNet-50, pruned ResNet-50, VGG-16, pruned VGG-16 and MobileNetV2, respectively, with higher accuracy of $2.83\%\uparrow$, $2.79\%\uparrow$, $3.57\%\uparrow$, $3.63\%\uparrow$ and $0.96\%\uparrow$. 
Additionally, through more aggressive compression, we introduce StriaNet-M1. Maintaining superior accuracy compared to the baseline, StriaNet-M1 achieves an increased speedup of $4.74\times$, $1.98\times$, $6.01\times$, $1.81\times$ and $2.41\times$.

For the CIFAR-10 dataset, shown in Table~\ref{ACC_Tiny_ImageNet_Cifar} and Figure~\ref{Flops_HE_tiny_cifar}, we compare StriaNet with the original ResNet-50, VGG-16, pruned VGG-16 and a CIFAR-10-specific architecture referred to as ResNet-56-CIFAR~\cite{he2016deep}. First, we introduce StriaNet-S2, achieved by further simplifying the StriaNet-M Series. StriaNet-S2 also demonstrates excellent performance, achieving a $6.1\times$, $3.65\times$, $1.27\times$ and $1.63\times$ speedup compared to ResNet-50, VGG-16, pruned VGG-16 and ResNet-56-CIFAR, respectively, with higher accuracy of $1.64\%\uparrow$, $0.64\%\uparrow$, $0.73\%\uparrow$ and $2.11\%\uparrow$. Additionally, with more aggressive compression, we introduce StriaNet-S1. With higher accuracy, StriaNet-S1 achieves an even higher $9.24\times$, $5.52\times$, $1.92\times$ and $2.81\times$ speedup.

\subsection{Detailed Training Setup}
\label{Training_Setup}

\vspace*{-0.1in}
\begin{table}[h]
\centering
\label{tab:my-table}
\resizebox{85mm}{22mm}{
\begin{tabular}{lll}
\Xhline{1.0pt}
\multicolumn{3}{c}{\textbf{General Setup}}                                                         \\ 
\multicolumn{1}{l}{$\cdot$Pytorch-v1.12.1} & \multicolumn{1}{l}{$\cdot$Optimizer: SGD}     & $\cdot$Seed: int(2)    \\ 
\multicolumn{1}{l}{$\cdot$Momentum: 0.9}     & \multicolumn{1}{l}{$\cdot$Weight decay: 5e-4} &                 \\ \Xhline{0.5pt}
\multicolumn{3}{c}{\textbf{StriaNet-Large on ImageNet}}                                            \\ 
\multicolumn{2}{l}{$\cdot$Learning rate: 0.01, decayed by 0.96 per epoch}              & $\cdot$Batch size: 128 \\ 
\multicolumn{2}{l}{$\cdot$Training epochs: 200 + 50 (optional)}                        &                 \\ \Xhline{0.5pt}
\multicolumn{3}{c}{\textbf{StriaNet-Mid on Tiny ImageNet}}                                         \\ 
\multicolumn{2}{l}{$\cdot$Learning rate: 0.01, decayed by 0.1 per 100 epochs}            & $\cdot$Batch size: 64  \\ 
\multicolumn{2}{l}{$\cdot$Training epochs: 120}                                        &                 \\ 
\Xhline{0.8pt}
\multicolumn{3}{c}{\textbf{StriaNet-Small on CIFAR-10}}                                            \\ 
\multicolumn{2}{l}{$\cdot$Learning rate: 0.01, decayed by 0.1 per 100 epochs}            & $\cdot$Batch size: 64  \\ 
\multicolumn{2}{l}{$\cdot$Training epochs: 250 + 50 (optional)}                        &                 \\ 
\Xhline{1.0pt}
\end{tabular}}
\end{table}

We train StriaNet with the ImageNet, Tiny ImageNet, and CIFAR-10 dataset using the specified setup shown above. The DistributedDataParallel (DDP) module~\cite{DDP} of PyTorch is employed for training with distributed data parallelism to accelerate the training process. When training with multiple GPUs (e.g., $N$ GPUs), the mini-batch size should be set as the actual batch size divided by $N$. We utilize SEED function to ensure the reproducibility of experimental results. However, disabling SEED can speed up the training process with minimal to no accuracy drop if reproducibility is not a concern.

Another observed situation is that when swapping between Windows and Linux OS with different versions of PyTorch, the trained model yields accuracy with very minor differences—some higher, some lower. This variation may be attributed to differences in the random mechanisms between PyTorch versions. For fairness, we selected the lowest accuracy to represent in this paper.

The trained models and codes are available as attachments to this paper. The trained models include StriaNet-L1 to StriaNet-L4 for ImageNet, StriaNet-M1 and StriaNet-M2 for Tiny ImageNet, and StriaNet-S1 and StriaNet-S2 for CIFAR-10.
All evaluations are provided through ``\textit{StriaNet\_$\times$\_eval.py}'' and a training code sample (StriaNet-L3, ImageNet, top-1 Acc: 76.33\%) is available via ``\textit{Training\_Sample\_StriaNet\_L3.py}''. The training adopts the \textit{``DistributedDataParallel"} for multi-GPU acceleration, and it can be configured and conducted by the command:

\vspace*{0.05in}
\noindent \textit{``OMP\_NUM\_THREADS=2 torchrun --nproc\_per\_node=3 --nnodes=1 Training\_Sample\_StriaNet\_L3.py"}.


\end{document}